\documentclass[a4paper,aps,reprint,onecolumn,notitlepage,superscriptaddress,nofootinbib]{revtex4-1}
\usepackage[utf8]{inputenc}
\usepackage[T1]{fontenc}

\usepackage{graphicx}
\usepackage{amsmath}
\usepackage{mathrsfs}  
\usepackage{braket}
\usepackage{amsfonts} 
\usepackage{amssymb} 
\usepackage{booktabs}
\usepackage{csquotes}
\usepackage[left=3cm,right=3cm,top=3cm,bottom=3cm]{geometry}
\usepackage{capt-of}
\usepackage[dvipsnames]{xcolor}

\usepackage[toc,title]{appendix}

\usepackage{siunitx}

\definecolor{vert}{rgb}{0.3,0.7,0.3}
\definecolor{bleu}{rgb}{0.3,0,1}
\definecolor{violet}{rgb}{0.2,0,0.7}

\definecolor{ct_black}{HTML}{000000}
\definecolor{ct_orange}{HTML}{ED872D}
\definecolor{ct_purple}{HTML}{7A68A6}
\definecolor{ct_blue}{HTML}{348ABD}
\definecolor{ct_turquoise}{HTML}{188487}
\definecolor{ct_red}{HTML}{E32636}
\definecolor{ct_pink}{HTML}{CF4457}
\definecolor{ct_green}{HTML}{467821}

\usepackage[breaklinks=true,colorlinks=true,linkcolor=ct_green,urlcolor=ct_green,citecolor=ct_green]{hyperref}
\usepackage{bookmark}

\def\tr{\ensuremath{\operatorname{tr}}}

\def\deg{\ensuremath{\operatorname{deg}}}
\newcommand{\ee}{{\rm e}}
\newcommand{\ii}{{\rm i}}
\newcommand{\dd}{{\rm d}}

\newcommand{\scat}{\mathcal{S}}

\newcommand{\ZZ}{\mathbb{Z}} 
\newcommand{\BZ}{\rm{BZ}}

\newcommand{\rot}{\mathscr{Z}}

\def\NN{\ensuremath{\mathbb{N}}}
\def\RR{\ensuremath{\mathbb{R}}}

\def\ZZ{\ensuremath{\mathbb{Z}}}

\def\Id{\ensuremath{\text{Id}}}
\def\UF{\ensuremath{U_\text{F}}}

\newcommand{\cc}{C_1}

\newcommand{\strong}[1]{\textbf{#1}}

\begin{document}

\title{Phase rotation symmetry \\ and the topology of oriented scattering networks}

\author{Pierre Delplace}
\email{Electronic address: pierre.delplace@ens-lyon.fr}
\affiliation{Univ Lyon, ENS de Lyon, Univ Claude Bernard Lyon 1, CNRS, Laboratoire de Physique, F-69342 Lyon, France}

\author{Michel Fruchart}
\affiliation{Univ Lyon, ENS de Lyon, Univ Claude Bernard Lyon 1, CNRS, Laboratoire de Physique, F-69342 Lyon, France}
\affiliation{Instituut-Lorentz, Universiteit Leiden, Leiden 2300 RA, The Netherlands}
\author{Clément Tauber}
\affiliation{Univ Lyon, ENS de Lyon, Univ Claude Bernard Lyon 1, CNRS, Laboratoire de Physique, F-69342 Lyon, France}
\affiliation{Dipartimento di Matematica, \enquote{La Sapienza} Università di Roma, Roma, Italy}

\date{\today}

\begin{abstract}
We investigate the topological properties of dynamical states evolving on periodic oriented graphs. This evolution, which encodes the scattering processes occurring at the nodes of the graph, is described by a single-step global operator, in the spirit of the Ho-Chalker model. When the successive scattering events follow a cyclic sequence, the corresponding scattering network can be equivalently described by a discrete time-periodic unitary evolution, in line with Floquet systems.

Such systems may present anomalous topological phases where all the first Chern numbers are vanishing, but where protected edge states appear in a finite geometry. To investigate the origin of such anomalous phases, we introduce the phase rotation symmetry, a generalization of usual symmetries which only occurs in unitary systems (as opposed to Hamiltonian systems). Equipped with this new tool, we explore a possible explanation of the pervasiveness of anomalous phases in scattering network models, and we define bulk topological invariants suited to both equivalent descriptions of the network model, which fully capture the topology of the system. We finally show that the two invariants coincide, again through a phase rotation symmetry arising from the particular structure of the network model.
\end{abstract}

\maketitle

\section{Introduction}

Topological insulators are remarkable materials where the particular topology of the bulk states leads to protected degrees of freedom with exceptional properties at the boundary of the system. For example, such edge states may provide a unidirectional propagation of  waves, and are robust against various perturbations. 
In this context, periodically driven (Floquet) dynamical systems have been shown to exhibit specific anomalous topological properties with no equivalent in equilibrium systems~\cite{KitagawaPRB2010, Rudner2013}. 
This anomalous behavior manifests itself by the existence of boundary states in finite geometry despite the vanishing of the topological index which usually accounts for all topological properties in equilibrium systems. More precisely, the first Chern number associated with the bands of the Bloch Hamiltonian that effectively describes the stroboscopic dynamics vanishes in this case.
The existence of these anomalous boundary states can instead be associated with a topological property of the full bulk evolution operator $U(t)$, which, unlike the effective Hamiltonian, accounts for the entire evolution at all times during one driving period~\cite{Rudner2013}.

This behavior can be generalized to a more general class of time-dependent dynamical systems. For linear systems, the evolution operator is generated by the Hamiltonian $H(t)$ of the system through an equation of motion $\ii  \partial_t U(t) = H(t) U(t)$ with initial condition $U(0) = \Id$,
which is formally solved by the time-ordered exponential
\begin{equation}
U(t) = \lim_{N \to \infty} \ee^{-\ii t/N \, H(N t/N)} \cdots \ee^{-\ii t/N \, H(n t/N)} \cdots \ee^{-\ii t/N \, H(t/N)}.
\label{eq:def_U}
\end{equation}
Namely, $U(t)$ results from an infinite product of infinitesimal free evolutions governed by instantaneous Hamiltonians $H(n t/N)$.
As the Hamiltonians at different times generically do not commute, the evolution operator $U(t)$ can be cumbersome to manipulate. 

However, it is often convenient to alternatively consider evolutions composed of a \emph{finite} sequence of \emph{step operations} described by unitary step operators $U_n$, so that after $N$ operations the evolution operator reads 
\begin{equation}
U = U_N U_{N-1}\dots U_1 \, .
\label{eq:Usequence}
\end{equation}
Such a stepwise dynamics suitably describes the effective discrete-time evolution of various experimental systems such as, in two dimensions, arrays of evanescently coupled optical waveguides with sufficiently sharp bending~\cite{Maczewsky2017,Mukherjee2017} and atomic discrete-time quantum walks, where the operators $U_n$ may consist of coin or shift operations applied to a spin-$1/2$ quantum state trapped in an optical lattice~\cite{Groth2016}.

Periodically driven systems include both evolutions generated by a time-periodic Hamiltonian $H(t)=H(t+T)$ and stepwise evolutions where the sequence of operations is repeated periodically. In both cases, the \emph{Floquet operator} of the evolution can be defined, respectively by $\UF = U(T)$ and by $\UF = U$. Despite their lack of explicit time dependence, stepwise evolutions were predicted to host anomalous topological chiral edge states in two dimensions, showing that the sequence structure~\eqref{eq:Usequence} is enough to engineer such topological phases~\cite{KitagawaPRB2010,  Rudner2013, KitagawaPRA2010, Liang2013, Pasek2014, Tauber2015, Asboth2015}.

\smallbreak

\begin{figure}[h!]
	\centering
\includegraphics{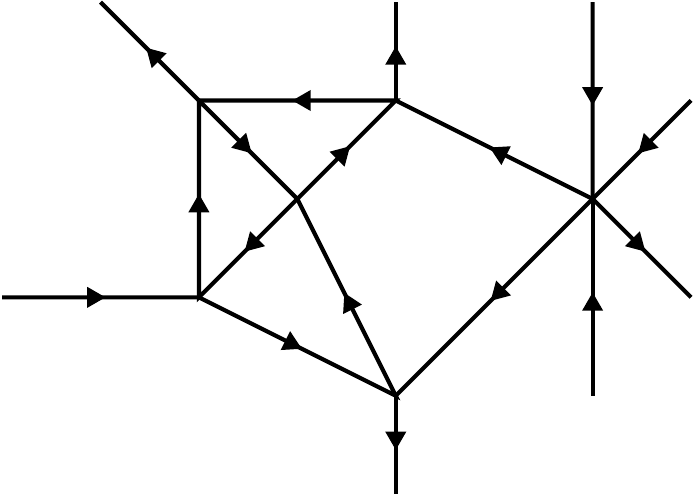}
	\caption{\label{fig:oriented}\strong{Example of an oriented network.} The direction of propagation along the links is represented by an arrow. The nodes represent the unitary scattering events between incoming and outgoing amplitudes.
		Due to the unitarity of scattering events, the number of incoming links is equal to the number of outgoing links at each node.
	}
\end{figure}

An important physical example was revealed by Liang, Pasek and Chong~\cite{Liang2013, Pasek2014} who described spatially periodic arrays of coupled photonic resonators in terms of unitary scattering matrices that locally encode the transmission and reflection coefficients of the optical signal between resonators, in order to go beyond the effective tight-binding description. Within this framework, the system can be seen as an \emph{oriented scattering network} similar to that introduced by Chalker and Coddington to describe the Hall plateau transition~\cite{Chalker1988,Ho1996}, which consists in links over which a directed current flows in one direction connecting nodes where incoming currents are scattered into outgoing currents, as represented in figure~\ref{fig:oriented}. 

Notably, the unidirectionality of the links plays a role similar to that of time as it forces the currents to cross the nodes in a given order that is fixed by the connectivity of the network. This behavior can originate from various physical mechanisms that explicitly break time-reversal symmetry, such as a perpendicular magnetic field like in the original Chalker-Coddington model~\cite{Chalker1988,Ho1996} or a flow of the propagation medium like in the array of acoustic circulators recently proposed by Khanikaev \emph{et al.}~\cite{Khanikaev2015} and Souslov \emph{et al.}~\cite{Souslov2016}. When time-reversal symmetry is preserved, as it is in most photonic systems, it is fair to use similar one-way oriented networks to describe one of the two \enquote{spin} copies of the system, provided that certain spin-flip processes can be neglected~\cite{Liang2013,Hu2015,Gao2016,Hafezi11}.
Due to this particular resurgence of an effective time, a fruitful analogy between scattering networks and Floquet dynamics was envisioned~\cite{Klesse1999,Janssen1999,Pasek2014}, an important consequence of which is the discovery of anomalous chiral edge states in such systems, while there is remarkably no external periodic driving as it would be in a Floquet system. The efficiency of this approach motivated two recent microwave experiments that probed the existence of these anomalous topological edge states~\cite{Hu2015, Gao2016}.

\bigbreak
Despite the accumulation of theoretical and experimental results on such systems, several questions remain open. First, the entire network is described by a unitary scattering matrix, the Ho-Chalker evolution operator~\cite{Ho1996}, which takes into account all the scattering events at the same time. In this picture, there is no Floquet dynamics, and the relation between both descriptions is not clear. A second issue is that even in the Floquet picture, a bulk topological characterization of network models is not available. The question of the characterization of the bulk topology of such systems is particularly crucial in the case of anomalous phases, where the first Chern numbers of the bands all vanish. Moreover, the way to engineer such phases remains an open question. Generically, bands of a two-dimensional gapped system where time-reversal symmetry is broken have a non-vanishing first Chern number. We therefore expect that an additional mechanism imposes their vanishing in certain conditions.

To answer this set of questions, we introduce in section~\ref{sec:Chern} a new symmetry specific to unitary systems, the \emph{phase rotation symmetry}, and show how this property constrains the value of the first Chern numbers associated to the spectral projectors of a gapped unitary operator.
In particular, a strong version of the phase rotation symmetry ensures the vanishing of first Chern numbers, a necessary condition to obtain anomalous topological phases. 

In oriented scattering networks, this phase rotation symmetry subtly enters at two different levels. First, it relates the evolution operator of certain networks to that of a Floquet-like system and allows for the definition of topological invariants. More precisely, a particular class of \emph{cyclic oriented networks} is introduced in section~\ref{sec:oriented_networks}, where the particular structure and connectivity of the scattering network \emph{constrains} its evolution operator to possess a particular phase rotation symmetry, which we call a structure constraint. This observation leads to several important results as it enables us to understand the structure of the evolution operator spectrum of the network model. Due to this insight, we are able to directly define a bulk topological invariant characterizing the system. The structure constraint also enables us to explicit the relationship between (cyclic) oriented network models and Floquet stepwise dynamics, and to define another bulk topological invariant for such dynamics. Indeed, both topological invariants are related and equivalent, as we finally show in section~\ref{sec:topo}.

The second role of the phase rotation symmetry in scattering networks is to provide an interpretation of the vanishing of first Chern numbers which is found in specific networks~\cite{Liang2013,Pasek2014}. At particular points of the phase diagram, another phase rotation symmetry, stronger than the structure constraint, may exist and enforce this vanishing. This allows us to propose a qualitative way to identify, in real space, whether a given oriented network may exhibit a vanishing first Chern number phase, which is developed in section~\ref{sec:anomalous_loop_configurations}.

\section{Unitary evolutions and the phase rotation symmetry}
\label{sec:Chern}

\subsection{Unitary evolutions and their phase spectra}

We consider systems described by a unitary evolution operator $U(t)$. This evolution may be derived from the microscopic description of the system, or rather be an effective description of the relevant degrees of freedom. We focus on situations where it is sensible to concentrate on the evolution operator $U=U(T)$ after some finite time $T$. Time-periodic dynamics provide the most common example of such a situation, as the evolution operator after one period $U(T)$ (here called the Floquet operator) describes the evolution of the system on long time scales. As we shall see in the next section, there are other cases where such a description is relevant ; this is in particular the case of oriented scattering networks, the study of which constitute the bulk of section~\ref{sec:oriented_networks}.

In a crystal, discrete space periodicity enables to block-diagonalize the evolution operator into a family of \emph{Bloch evolution operators} $U(k)$ which are finite matrices, and are labeled by a quasi-momentum $k$ living on a $d$-dimensional torus called the Brillouin torus (we will only consider the case $d=2$ here). The spectrum of the evolution operator $U$ is called its \strong{phase spectrum}. Its eigenstates $\ket{\psi_n(k)}$ satisfy the eigenvalue equation
\begin{equation}
U(k) \ket{\psi_n(k)} = \ee^{-\ii \varepsilon_n(k)} \ket{\psi_n(k)}
\label{eq:U}
\end{equation}
where the eigenphases $\varepsilon_n(k)$, which constitute the phase spectrum, are confined on the unit circle in the complex plane. The minus sign in~\eqref{eq:U} is arbitrary; it is chosen here for the analogy between $U$ and the evolution operator. 

\begin{figure}[htb]
  \centering
\includegraphics{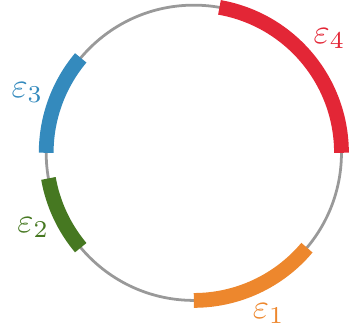}
  \caption{\label{fig:phase_spectrum}\strong{Phase spectrum.} Illustration of a phase spectrum with four bands and four gaps.}
\end{figure}

Generically, the phase spectrum displays \strong{phase bands} separated from each other by \strong{phase gaps}, as illustrated in figure~\ref{fig:phase_spectrum}. Each band corresponds to a family of orthogonal projectors $k \mapsto P(k)$, which describe the spectral projector on the corresponding arc in the unit circle.

\subsection{The phase rotation symmetry \label{subsec:PRS}}

\begin{figure}[htb]
  \centering
 \includegraphics{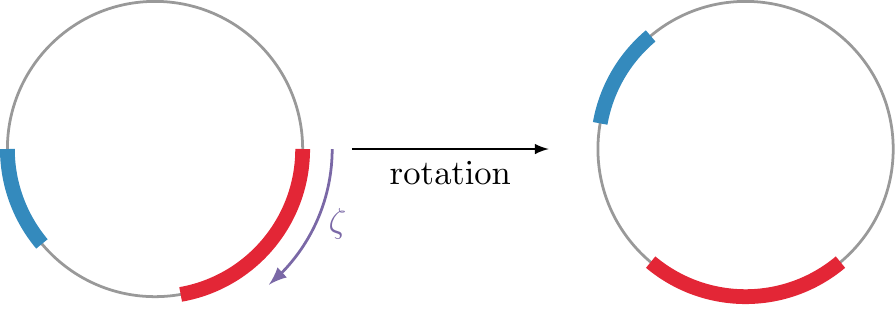}
  \caption{\label{fig:phase_eigenvalue_rotation}\strong{The phase rotation.} On the level of the spectrum, a phase rotation of angle $\zeta$ maps eigenvalues $\ee^{-\ii \varepsilon}$ to $\ee^{-\ii (\varepsilon+\zeta)}$.}
\end{figure}

Unitary systems share the particularity to have a periodic spectrum. This allows us to consider a rotation of those spectra by an angle $\zeta$, corresponding to the transformation $\ee^{-\ii \varepsilon} \to \ee^{-\ii (\varepsilon + \zeta)}$, as depicted in figure~\ref{fig:phase_eigenvalue_rotation}.

We consider situations where the phase spectrum is invariant under such a phase rotation.
Although a symmetry of the phase spectrum can be accidental, this situation is not typical and instead we consider situations where the invariance of the phase spectrum under such a phase rotation is associated to a \strong{phase rotation symmetry} of the form 
\begin{equation}
  \rot U \rot^{-1} = \ee^{\ii \zeta} U
  \label{eq:symmetry_def}
\end{equation}
where $\rot$ is a unitary \strong{phase rotation operator} acting on the states of the Hilbert space.

\noindent The phase rotation symmetry~\eqref{eq:symmetry_def} is the evidence of a redundancy in the description of the system. Indeed, if $\ket{\psi}$ is an eigenstate of $U$ with eigenvalue $\ee^{-\ii \varepsilon}$, then $\rot \ket{\psi}$ is also an eigenstate of $U$, with the eigenvalue $\ee^{-\ii (\varepsilon+\zeta)}$ ; more generally, $\rot^m \ket{\psi}$ (with $m$ an integer) is an eigenstate of $U$ with eigenvalue $\ee^{-\ii(\varepsilon+m\zeta)}$.

Crucially, such a \enquote{symmetry} has no equivalent in Hamiltonian systems, as it would correspond to an unphysical energy translation $E \to E + \Delta E$. In contrast, it can arise in \enquote{unitary systems} as the phase spectrum lies on a circle.

When $\zeta/2\pi$ is irrational, the irrational rotation of the Floquet spectrum ensures that it is fully gapless. 
On the other hand, when $\zeta/2\pi = m/M$ is a rational, where $m/M$ is an irreducible fraction, a phase is mapped to itself applying the phase rotation $M$ times. Phases being defined modulo $2\pi$ it is sufficient to consider $0 \leq \zeta \leq 2\pi$, so we can set $m=1$ without loss of generality. 
As we are interested in gapped unitary operators, we will focus on cases where $\zeta = 2\pi/M$  where $M$ is an integer. The phase rotation symmetry~\eqref{eq:symmetry_def} then reads
\begin{equation}
\rot U \rot^{-1} =  \ee^{\ii 2\pi/M} U \ .
\label{eq:constraint}
\end{equation}
In practice, it is more convenient to use the Bloch version of this symmetry. Assuming that the operator $\rot$ is local in space (i.e. it does not couple different unit cells),~\eqref{eq:constraint} straightforwardly translates as $\rot U(k) \rot^{-1} =  \ee^{\ii 2\pi/M} U(k)$ where $U(k)$ is the Fourier transform of $U$.
As the variable $k$ is not affected by the phase rotation, we will omit it when the meaning is clear.

\begin{figure}[htb]
  \centering
 \includegraphics{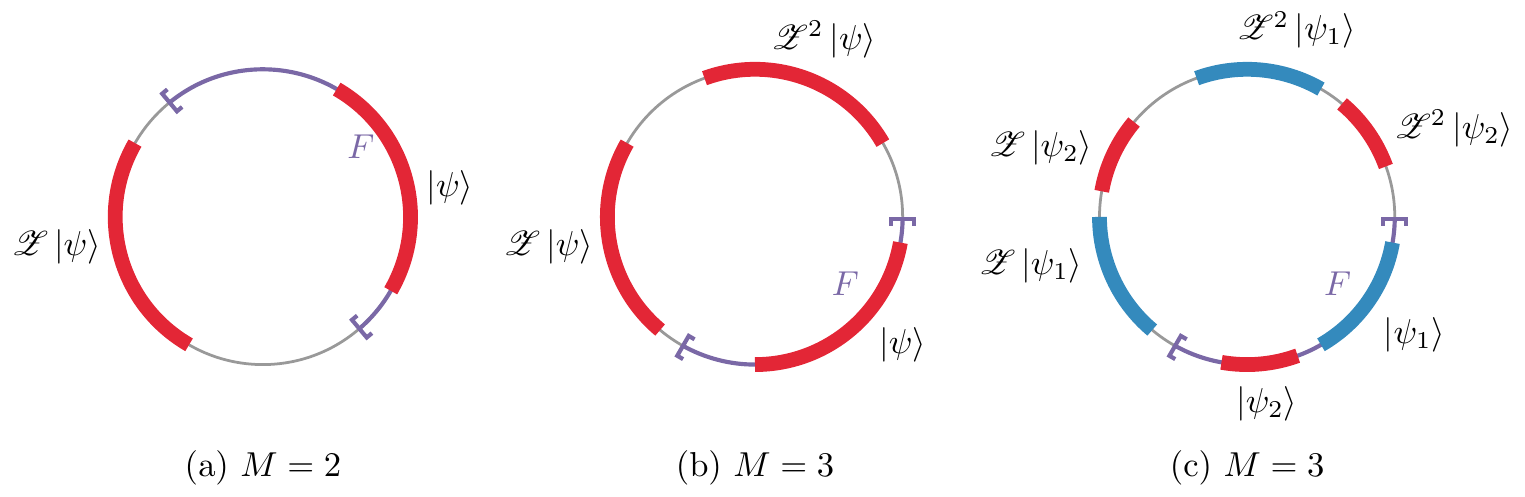}
   \caption{\label{Floquet_eigenvalue_rotation_invariant_spectra}\strong{Examples of phase rotation invariant spectra.} The spectra are invariant under a rotation of $2\pi/M$, with (a) $M=2$ and (b,c) $M=3$. In all cases, a fundamental domain $F$ for the symmetry can be chosen. For our purposes, the most convenient choice is an interval of length $2\pi/M$ with both ends lying in a spectral gap. In each case, a possible choice of fundamental domain is represented in purple.}
\end{figure}

Let us assume that $U$ has a gap around $\ee^{-\ii \eta}$. Then, due to the phase rotation symmetry~\eqref{eq:constraint}, there is also a gap around  $\ee^{-\ii (\eta + 2\pi/M)}$. A \strong{fundamental domain} $F$ for the phase rotation symmetry is then defined by the interval between these two values, so that it represents the shorter arc that links $\ee^{-\ii \eta}$ and $\ee^{-\ii (\eta + 2\pi/M)}$ on the unit circle (see figure~\ref{Floquet_eigenvalue_rotation_invariant_spectra}). The fundamental domain $F$ plays a role similar to that of a unit cell: starting from the part of the spectrum over the arc $F$, the whole spectrum is recovered by $M$ successive applications of the phase rotation of an angle $2\pi/M$ (for eigenvalues) and of the unitary operator $\rot$ (for eigenvectors) as illustrated in figure~\ref{Floquet_eigenvalue_rotation_invariant_spectra}. 

Phase rotation symmetry allows one to reduce the description of the system by removing its redundancy, essentially by keeping only the eigenstates in one fundamental domain.
As an example, this reduction procedure will be carried out explicitly in the case where $\rot^M = \Id$ during the study of oriented scattering networks in section~\ref{sec:oriented_networks}, and it will allow us to account for the topological properties of such systems.

\bigbreak

Notice that~\eqref{eq:constraint} implies that
\begin{equation}
\rot U^M \rot^{-1} =  U^M,
\label{eq:sym_recovered}
\end{equation}
which has the usual form of an \emph{actual symmetry} (with an equivalent in Hamiltonian systems) for~$U^M$. Equation~\eqref{eq:sym_recovered} means that the system recovers a symmetry represented by the operator $\rot$ after $M$ successive identical evolutions.

\smallbreak

Besides, the $M^{\text{th}}$ power of the phase rotation operator is also a symmetry of $U$, as
\begin{equation}
	\rot^{M} U \rot^{-M} = U.
\end{equation}
In general, this symmetry can be arbitrary.
When $\rot^M$ is scalar and $U$ is gapped, the phase rotation operator assumes the particular form
\begin{equation}
	\label{eq:standard_prs}
	\rot \simeq \text{diag}(1,\ee^{\ii 2\pi/M}, \ee^{\ii 2\pi \times 2/M}, \dots , \ee^{\ii 2 \pi (M-1)/M}) \otimes \Id = \rot_0
\end{equation}
in an adequate basis, which emphasizes its cyclic behavior (see appendix \ref{app:standard_PRS}).

\subsection{Topological states and the phase rotation symmetry}

As we have seen, phase rotation symmetry enables to reduce the degrees of freedom in the description of the system.
Another important consequence of this symmetry is to impose particular constraints on the topological properties of the system. As we shall see, a crucial consequence of the phase rotation symmetry is that the spectral projector over one fundamental domain has a vanishing first Chern number. 

\bigbreak

For concreteness, we focus on two dimensional crystals in the following. Each band of the evolution operator $U$ carries a first Chern number, which is computed from the projector family $k \mapsto P(k)$ as
\begin{equation}
	\cc(P) = \frac{\ii}{2\pi} \int \tr P \dd P \wedge \dd P \, .
 \label{eq:def_chern}
\end{equation}
Let us recall two important properties of the first Chern number which will be useful later. First, it is invariant under conjugation by a constant unitary operator $\mathcal{U}$,
\begin{equation}
	\label{chern_conjugation}
	\cc(\mathcal{U} P \mathcal{U}^{-1}) = \cc(P).
\end{equation}
Moreover, it is additive: if $P$ and $Q$ are mutually orthogonal projector families (so $P Q = 0 = Q P$), then
\begin{equation}
	\label{chern_additivity}
	\cc(P+Q) = \cc(P) + \cc(Q).
\end{equation}

A nonvanishing first Chern number signals a nontrivial bulk topology of the system, which manifests itself in the appearance of robust chiral edge states at the boundary of a finite sample. When $U$ corresponds to a time-independent Hamiltonian evolution, the first Chern numbers fully characterize the bulk topological properties of the system (at least in the Altland-Zirnbauer symmetry class A). In general, however, this is not the case: there are the so-called \emph{anomalous topological phases} which display a nontrivial topology despite having vanishing first Chern numbers~\cite{KitagawaPRB2010,Rudner2013}. Such topological properties are instead captured by taking into account the full time-dependent evolution in the bulk~\cite{Rudner2013}, and not only the bulk evolution operator after a finite amount of time. 


\subsubsection{Consequences of the phase rotation symmetry}
\label{sec:consequences}

Let us denote by $\Pi$ the spectral projector on states with eigenvalues $\ee^{-\ii \varepsilon} \in F$, i.e. on a fundamental domain.
The spectral projector $\Pi_m$ on the $m$-th rotated fundamental domain $\ee^{-\ii 2\pi m /M} F$ is then obtained by the action of $\rot$ as $\Pi_m=\rot^{m} \Pi \rot^{-m}$. Due to the invariance of the first Chern number under conjugation by a constant unitary operator~\eqref{chern_conjugation}, all the rotated fundamental domains have the same first Chern number
\begin{equation}
	\cc(\Pi_m) = \cc(\rot^m \Pi \rot^{-m}) = \cc(\Pi).
\end{equation}
Second, as these projectors sum to identity 
\begin{equation}
	\sum_{m=0}^{M-1} \rot^{m} \Pi \rot^{-m} = \Id
\end{equation}
and due to the additivity of the first Chern number~\eqref{chern_additivity}, we infer that
\begin{equation}
	\label{vanishing_chern}
	 \cc(\Pi) = 0.
\end{equation}
As a consequence, the first Chern number of the spectral projector on any rotated fundamental domain vanishes. This is one of the main results of this paper.

\bigbreak

In general, the projector $\Pi$ on a fundamental domain $F$ of the phase rotation symmetry does not correspond to a single band, as there may be phase gaps inside of $F$ (see figure~\ref{Floquet_eigenvalue_rotation_invariant_spectra}). 
In the particular situation where $\Pi$ \emph{does} correspond to a single band\footnote{A single band does not necessarily correspond to a single \emph{state}. The projector $\Pi$ may have a rank higher than one, provided that the corresponding eigenstates of $U$ are degenerate (at least at some point of the Brillouin torus).}, we say that the evolution operator is endowed with a \strong{strong phase rotation symmetry}.
It follows from the previous discussion that in this situation, the first Chern numbers of each band in the spectrum of $U(T)$ vanish. As a consequence, \emph{the corresponding phase is either topologically trivial or anomalous}. This observation is particularly interesting as it provides an explanation to the prevalence of anomalous topological states in certain contexts. When time-reversal symmetry is broken, we typically expect the appearance of nonvanishing first Chern numbers, at least when the corresponding phase does not include a time-reversal invariant point. However, there are systems where only anomalous phases appear (a concrete example is discussed in section~\ref{sec:L_lattice}): this surprising behavior is explained by the existence of a phase rotation symmetry (at least at particular points of the phase diagram) which prevents nonvanishing first Chern numbers from appearing, despite the breaking of time-reversal symmetry.

\section{Oriented scattering network models and periodic sequences of steps}
\label{sec:oriented_networks}

\subsection{Introduction}

The propagation of waves in a time-reversal breaking metamaterial can be described by an \strong{oriented scattering network} composed of unitary scattering matrices (the nodes) connected to each other by oriented links. The dynamics of waves in the network model are then described by a unitary evolution operator which contains all the vertex scattering matrices as well as the connectivity of the network.

Oriented networks models were originally introduced by Chalker and Coddington to describe the Hall plateau transition~\cite{Chalker1988,Ho1996}. 
In a semi-classical picture, electronic wave packets in a disordered two-dimensional electron gas under strong magnetic field follow the equipotentials of the smooth disorder potential, in a direction fixed by the magnetic field. The quantum Hall transition essentially arises when the equipotentials of the disorder percolate; however, near the transition, the relevant equipotentials approach the saddle points of the disorder potential and become closer and closer. Hence, wave packets can tunnel from an equipotential to another giving rise to \enquote{quantum percolation}~\cite{Chalker1988} (see also~\cite{Cardy2010} for a pedagogical introduction). This process is described by scattering matrices, one per saddle point, within the Chalker--Coddington model~\cite{Chalker1988} that distorts the equipotentials into a periodic square lattice of such scattering matrices connected by incoming and outgoing directed links, the so-called L-lattice. Remarkably, this oriented network model captures most of the essential features of the Hall plateau transition.
In the original model~\cite{Chalker1988}, random phases are added on each link to take into account the Aharanov--Bohm phase accumulated on the closed disorder equipotentials of various sizes. A fully space-periodic oriented network, without such random phases, was introduced by Ho and Chalker~\cite{Ho1996}, who showed that a Dirac equation emerged from an expansion of a discrete evolution operator of the scattering network model.

More recently, Liang, Pasek and Chong~\cite{Liang2013, Pasek2014} introduced a similar formalism to investigate the properties of an array of coupled photonic resonators beyond tight-binding descriptions. In such a system, the coupling between resonators is described by unitary scattering matrices that encode the transmission and reflection coefficients of the optical signal, rather than by an effective tight-binding Hamiltonian. The same formalism was also applied to sound waves in arrays of acoustic circulators by Khanikaev \emph{et al.}~\cite{Khanikaev2015}. In both situations, the light or sound waves in the system are described by a huge scattering matrix, which can be understood as the evolution operator of the system. Notably, robust chiral edge states appear in a finite system, precisely in the phase gap(s) of the bulk scattering matrix. This is not a surprise in light of the connection with the quantum Hall effect. What is more surprising is that Liang, Pasek and Chong unveiled that photonic arrays support anomalous topological states similar to that described by Rudner et al. in periodically driven systems~\cite{Rudner2013}, despite the lack of explicit time dependence of the system.

\bigbreak

The existence of such anomalous topological states appears to be a fundamental property of unitary systems, as it crucially depends on the periodicity of the phase spectrum; such a behavior may in principle emerge whenever a unitary description of the system is adopted~\cite{Tauber2015}. In contrast, they are not captured in an effective tight-binding description~\cite{Liang2013,Hafezi11}. 

As we have seen in Section~\ref{sec:consequences}, the pervasiveness of anomalous phases can be attributed to the existence of particular constraints, like a phase rotation symmetry. This is indeed the case in oriented scattering networks, where anomalous phases can be tracked down to the existence of particular \enquote{symmetric points} of the phase diagram where a phase rotation symmetry is present. As we shall see, such phase rotation symmetries can further be interpreted in terms of \emph{classical loop configurations} of the oriented network. This interpretation is particularly powerful as it allows one to \emph{design} anomalous phases in a straightforward way.

\bigbreak

In a potentially topological anomalous system,  first Chern numbers are not sufficient to distinguish the possible topological phases (as they are always zero), and more precise bulk invariants are required. Crucially, the unidirectionality of the links plays a similar role to that of time as it forces the wave packets to visit the vertices in a given order which is determined by the connectivity of the network. In the following, we define a class of scattering networks, \strong{cyclic oriented networks}, where it is possible to map the network model to a (stepwise) time-dependent system to study its properties. This mapping is allowed by the existence of a \strong{structure constraint} which encodes the particular connectivity of the cyclic oriented network. On the level of the evolution operator describing the entire scattering network, this constraint manifests itself as a phase rotation symmetry, which allows for the definition of bulk topological invariants that fully characterize the network model.

\subsection{Oriented scattering network models}

In general, oriented scattering network models consist of a directed graph, composed of a set of vertices (or nodes) representing scattering matrices, which are connected to each other by directed edges (or links) over which flows a directed current~\cite{Cardy2010}. At each vertex $v$, the number $b_v$ of incoming links is equal to the number of outgoing links to guarantee the unitarity of scattering events, which are described by a scattering matrix $S_v \in U(b_v)$, which relates the incoming amplitudes $c^{\text{in}}_e$ on each incoming edge $e$ to the outgoing amplitudes $c^{\text{out}}_f$ on each outgoing edge $f$ by
\begin{equation}
  c^{\text{out}}_f = (S_v)_{f e} \, c^{\text{in}}_e.
\end{equation}
Here, we will only consider spatially periodic graphs. There may be several scattering matrices in a unit cell, but for simplicity we will further assume that all scattering matrices have the same size $b$, i.e. that each vertex is connected to the same number of links. The most simple nontrivial situations is $b=2$, where matrices are $U(2)$ rotations, and it is usually possible to reduce any network model to this situation~\cite{Cardy2010,Cardy2005}. 

\smallbreak

While network models can be used in any space dimension, we shall focus on two-dimensional systems.
Waves in such a spatially periodic network are described by a unitary Bloch scattering matrix. In the bulk, Bloch reduction gives a matrix $S(k_x,k_y)$ from which one can hope to extract topological invariants. In a finite cylinder geometry, a bigger matrix $S_{\text{cylinder}}(k_y)$ (whose size depends on the height of the cylinder) describes both the bulk and the edge states of the finite system. In both cases, we obtain a periodic phase spectrum : as usually in topological systems, the bulk phase gaps host the chiral anomalous edge states that appear in a finite geometry.

\subsection{An archetypal example: the L-lattice}
\label{sec:L_lattice}

One of the simplest examples of oriented scattering networks is the L-lattice, which was introduced by Chalker and Coddington~\cite{Chalker1988}. We illustrate the main focal points of our analysis on this example, namely (i) the definition of bulk topological invariants that fully characterize the network model and (ii) the existence of special points of the phase diagram where a strong version of the phase rotation symmetry ensures the vanishing of the first Chern numbers, allowing only for anomalous topological phases.

\begin{figure}[h!]
\centering
\includegraphics{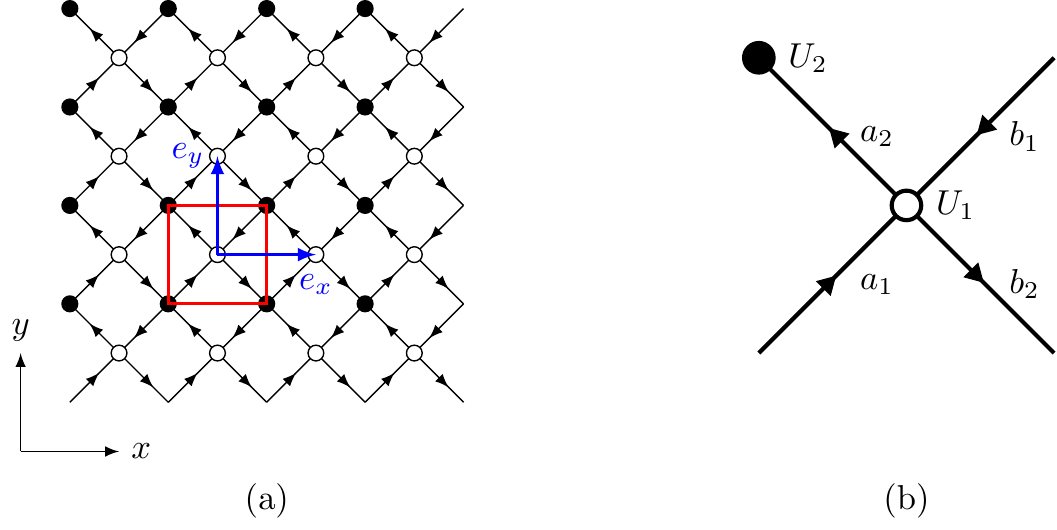}
\caption{\strong{L-lattice.} (a) The L-lattice as a square Bravais lattice with basis vectors $e_x$ and $e_y$, with its four inequivalent links and two inequivalent nodes. A unit cell is enhanced in red and detailed in (b).}
\label{fig:L-lattice}
\end{figure}

The L-lattice is an oriented network model on a square Bravais lattice with two inequivalent vertices and four inequivalent oriented links per unit cell (which somehow ressembles two L letters connected together). More precisely, the unit cell is composed of two vertices $U_1$ and $U_2$ and of four inequivalent oriented links $(a_1,b_1,a_2,b_2)$ connecting the vertices, as represented in figure~\ref{fig:L-lattice}. The unitary matrices $U_j \in U(2)$ encode how amplitudes on their two incoming links are scattered into their two outgoing links, as
\begin{equation}
  \begin{pmatrix}
    a_2(x,y,t+T) \\
    b_2(x,y,t+T)
  \end{pmatrix}
  =
  U_1 \,
  \begin{pmatrix}
    a_1(x,y,t) \\
    b_1(x,y,t)
  \end{pmatrix}
\end{equation}
and
\begin{equation}
  \begin{pmatrix}
    a_1(x,y+1,t+T) \\
    b_1(x-1,y,t+T)
  \end{pmatrix}
  =
  U_2 \,
  \begin{pmatrix}
    a_2(x,y,t) \\
    b_2(x-1,y+1,t)
  \end{pmatrix}.
\end{equation}
For simplicity, we choose the parametrization
 \begin{equation}
U_j= \begin{pmatrix}
\cos{\theta_j}& \sin{\theta_j} \\
-\sin{\theta_j} & \cos{\theta_j}
\end{pmatrix}
\end{equation}
of the vertex scattering matrices, where the parameters $\theta_j$ control the transmission and reflection at each vertex. Complex phases can be introduced but will not change the properties we discuss here, namely the existence of two distinct topological phases both with vanishing first Chern numbers~\cite{Liang2013}. Moreover, for convenience and to compare with~\cite{Liang2013}, we focus on the situation where both angles are controlled by a single parameter $\theta = \theta_2 = \pi/2 - \theta_1$. 

A state $\ket{\psi}$ of the system is given by a set of amplitudes $\{a_1(x,y)$, $b_1(x,y)$, $a_2(x,y)$, $b_2(x,y)\}$ for all positions $(x,y)$ in the square Bravais lattice. Following Ho and Chalker~\cite{Ho1996}, we consider the \emph{discrete evolution operator} $\scat$ that describes the evolution of a state $\ket{\psi}$ after its amplitude on each link has been scattered at the nodes of the network. In other words, this operator effectively describes the  scattering processes at all the nodes simultaneously. 

\smallbreak

When focusing on the stationary bulk states, we can assume translation invariance and Fourier transform both the stationary states and the evolution operator into their block-diagonal Bloch version. The Bloch version of the Ho-Chalker evolution operator reads
\begin{equation}
\label{eq:ho_chalker_Llattice}
 \scat(k) = 
\begin{pmatrix}
 0        & U_2(k) \\
 U_1(k)   & 0
\end{pmatrix} 
\end{equation}
in the Bloch basis $(a_1(k),b_1(k),a_2(k),b_2(k))$, where $k$ is in the two-dimensional Brillouin zone.
For the choice of unit cell shown in figure~\ref{fig:L-lattice}(b), the two unitary blocks are given by
\begin{equation}
U_1(k) = \begin{pmatrix}
\sin{\theta} & \cos{\theta} \\
-\cos{\theta} & \sin{\theta}
\end{pmatrix}
\qquad
\text{and}
\qquad
U_2(k) = \begin{pmatrix}
\cos{\theta}\, \ee^{-\ii k_y}& \sin{\theta}\,\ee^{-\ii k_x} \\
-\sin{\theta}\, \ee^{\ii k_x}& \cos{\theta}\, \ee^{\ii k_y}
\end{pmatrix}.
\end{equation}

The block-antidiagonal form~\eqref{eq:ho_chalker_Llattice} of the evolution operator is reminiscent of the \emph{cyclic structure} of the oriented network:  as $a_1$ and $b_1$ are oriented from $U_2$ to $U_1$, whereas $a_2$ and $b_2$ are oriented from $U_1$ to $U_2$, a wave packet traveling in the network will always encounter a succession $U_1 \to U_2 \to U_1 \to U_2 \to \cdots$ of nodes (and will never, for example, come across two successive $U_1$ nodes).
It is convenient to reframe this particular block-antidiagonal form in terms of the \strong{structure constraint}
\begin{equation}
  \label{eq:structure_constraint_L}
  D \scat(k) D^{-1} = - \scat(k)
  \qquad
  \text{where}
  \qquad
  D = \begin{pmatrix}
    \Id & 0 \\ 0 & - \Id
  \end{pmatrix}
\end{equation}
where $\Id$ is here the two-by-two identity matrix. We recognize a particular case of the phase rotation symmetry~\eqref{eq:constraint} with $\rot=D$ and $M=2$. As we shall see in section~\ref{sec:cyclic}, such a structure constraint can be generalized to a whole class of network models. (On first sight, this particular case may looks like a chiral symmetry, but this is not the case as $\scat$ is an evolution operator and not a Hamiltonian.)

\bigbreak

\begin{figure}[htb]
  \centering
\includegraphics{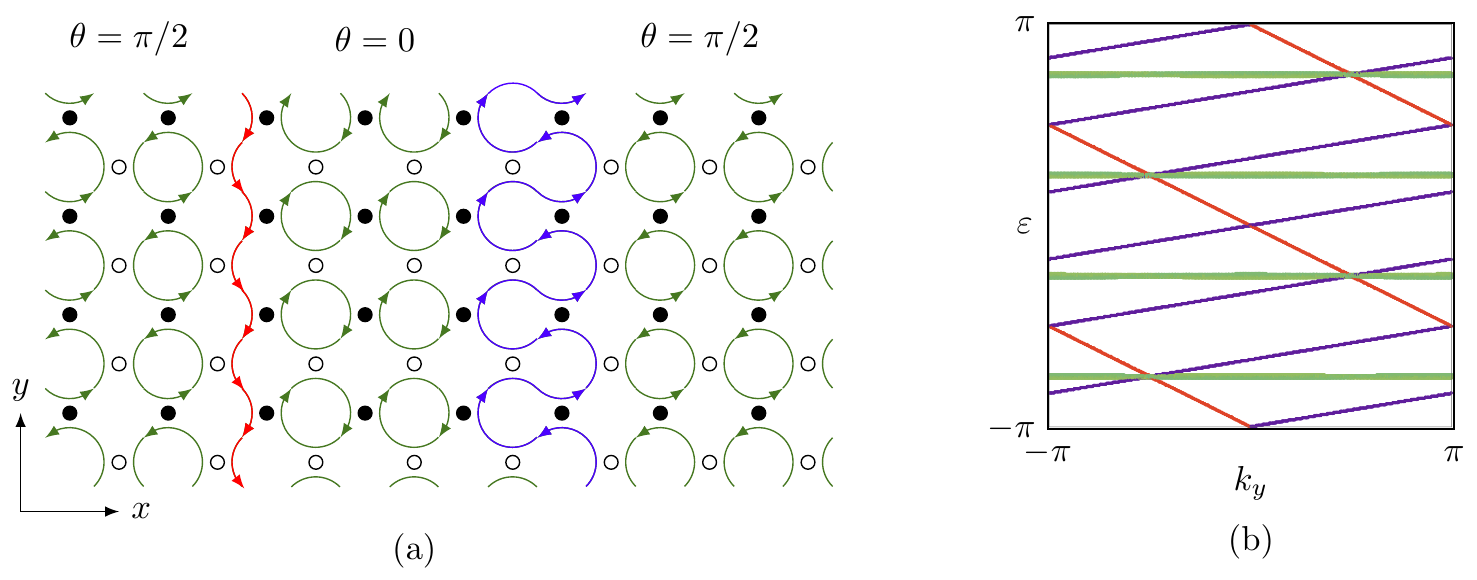}
  \caption{\label{fig:interface_L_lattice}\strong{Interfaces of the L-lattice.} 
  We consider interfaces between the two phases of the L-lattice in a cylinder geometry (the system has periodic boundary conditions in the $x$ direction and is infinite in the $y$ direction).  This allows one to (i) avoid potential ambiguities due to the relative character of the invariant and (ii) confirm that the existence of chiral edge states is indeed due to the bulk topology, and not merely from the oriented nature of the links.
  Remarkably, the two chiral edge states (one at each interface) are found to have different group velocities, which is consistent with the simple intuitive sketch in (a) where one of the two channels (in red) can flow easily rather than the other one (in blue) is forced to propagate in pilgrimage, resulting in a decreasing of its velocity along the $y$ axis compared to that of the other boundary state. 
  (a) Interfaces between two networks with respectively $\theta=0$ and $\theta=\pi/2$. The system in periodic in both direction and finite in the $x$ direction. At the two interfaces, edge states with different velocity, in sign and amplitude, arise. (b) Eigenvalues of the corresponding Ho-Chalker evolution operator with $\theta\approx 0$ and $\theta \approx \pi/2$ for clarity. Bulk states are represented in green, the fast boundary state in red and the slow boundary state in blue.
  The code used to compute the phase spectra and the topological invariants is available in Supplemental Materials at URL \url{https://arxiv.org/src/1612.05769/anc}.
  }
\end{figure}

The well-known phase diagram of the L-lattice~\cite{Ho1996,Liang2013} with respect to the parameter $\theta$ is represented in figure~\ref{fig:L-loops}. Due to the form of matrices $U_j$, it is $\pi$-periodic with respect to $\theta$, and we can restrict the discussion to a range of that length. The phase spectrum of $\scat(k)$ consists in four bands that touch at the critical value $\theta_c=\pi/4$, and this critical point separates two phases where the four bands are well-defined (i.e. separated by gaps), which we call phases I and II. Notably, such phases are topologically inequivalent, a smoking gun evidence of which is the existence of robust chiral edge states at an interface between them (see figure~\ref{fig:interface_L_lattice}).

\smallbreak 

Following a longstanding analogy between network models and Floquet stepwise evolutions~\cite{Klesse1999,Janssen1999}, Liang, Pasek and Chong~\cite{Liang2013,Pasek2014} studied the topology of network models by focusing on the Floquet operator $U_{\text{F}}(k)=U_2(k)U_1(k)$ which represents a \emph{sequence} of two steps, in contrast with the Ho-Chalker evolution operator $\scat(k)$ that accounts for the different scattering processes \emph{simultaneously}. 
The equivalence between both points of view is rooted into the existence of the structure constraint~\eqref{eq:structure_constraint_L}. Due to this phase rotation symmetry, the description of the system from the point of view of the Ho-Chalker evolution operator $\scat(k)$ is redundant, and its spectrum reduced to a fundamental domain is directly related to the (entire) spectrum of $U_{\text{F}}(k)$.
The structure constraint enables to define bulk invariants that characterize the network model: for each bulk gap $\ee^{-\ii \eta}$ of the Ho-Chalker evolution operator $k\mapsto \scat(k)$, there is a bulk invariant
\begin{equation}
  W_{\eta}^{\text{HC}}[\scat] \in \ZZ
\end{equation}
which essentially accounts for the number of edge states appearing in the bulk gap $\ee^{-\ii \eta}$ when an interface is considered. We defer the definition of such invariants to the section~\ref{sec:topo}, but we will now discuss their essential properties. The redundancy expressed by the phase rotation symmetry~\eqref{eq:structure_constraint_L} is translated at the level of such invariants by the identity
\begin{equation}
  W_{\eta}^{\text{HC}}[\scat] = W_{\eta+2\pi/M}^{\text{HC}}[\scat]
\end{equation}
where $M=2$ in the case of the L-lattice.

\smallbreak

Crucially, this invariant is \emph{relative} to a reference evolution which has to be chosen arbitrarily. For the unit cell in figure~\ref{fig:L-lattice}, we obtain $W_{0}^{\text{HC}}[\scat_{\text{I}}]=\num{1}$ and $W_{\pi/2}^{\text{HC}}[\scat_{\text{I}}]=\num{1}$ in phase I and $W_{0}^{\text{HC}}[\scat_{\text{II}}]=\num{0}$ and $W_{\pi/2}^{\text{HC}}[\scat_{\text{II}}]=\num{0}$ for phase II. A different choice of unit cell leads to different values for the invariants (see table~\ref{fig:L_lattice_relative_invariants} for an example, and section~\ref{sec:relative} for a more detailed discussion), yet the differences between invariants \emph{do not} depend on particular choices. Usually, only such differences carry a physical meaning; for example, their variation at an interface is expected to give the algebraic number of chiral edge states (counted with chirality) in the corresponding bulk gap. Particular physical situations may however naturally select only one unit cell.

\begin{figure}
\centering
\includegraphics{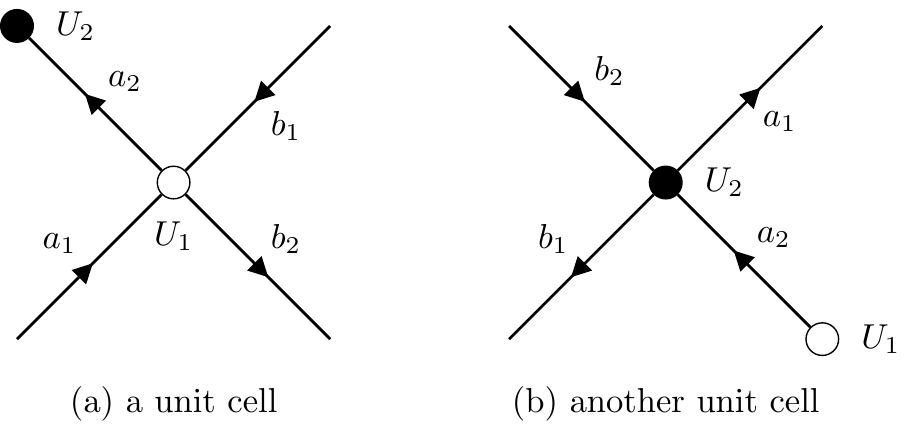}
\caption{\label{fig:L_lattice_different_unit_cells}\strong{Two possible unit cells of the L-lattice.}}
\end{figure}

\begin{table}[htb]
\centering
\begin{tabular}{lSS}
\toprule
unit cell & {(a)} & {(b)} \\
\midrule
$W_{0}^{\text{HC}}[\scat_{\text{I}}]$ & 1 & 0 \\
$W_{0}^{\text{HC}}[\scat_{\text{II}}]$ & 0 & -1 \\
$W_{0}^{\text{HC}}[\scat_{\text{I}}] - W_{0}^{\text{HC}}[\scat_{\text{II}}]$ & 1 & 1 \\
\midrule
$W_{\pi/2}^{\text{HC}}[\scat_{\text{I}}]$ & 1 & 0 \\
$W_{\pi/2}^{\text{HC}}[\scat_{\text{II}}]$ & 0 & -1 \\
$W_{\pi/2}^{\text{HC}}[\scat_{\text{I}}] - W_{\pi/2}^{\text{HC}}[\scat_{\text{II}}]$ & 1 & 1 \\
\bottomrule
\end{tabular}
\caption{\label{fig:L_lattice_relative_invariants}\strong{Relative invariants for the L-lattice.} The values of the invariants are given for two choices of a unit cell (a) and (b), for the two phases I and II, and for the two gaps $\eta=0$ and $\eta=\pi$. We observe that the values \emph{do not} coincide when the unit cell changes, but that the difference between two phases $W_{\eta}^{\text{HC}}[\scat_{\text{I}}] - W_{\eta}^{\text{HC}}[\scat_{\text{II}}]$ is invariant with respect to the choice of the unit cell, as it is expected for a physically observable quantity. The chosen unit cells are represented in figure~\ref{fig:L_lattice_different_unit_cells}, and a more detailed account of the choice of the reference evolution is explained in the general case, in section~\ref{sec:relative}. 
The code used to compute the phase spectra and the topological invariants is available in Supplemental Materials at URL \url{https://arxiv.org/src/1612.05769/anc}.}
\end{table}

\bigbreak

\subsubsection{Classical loop configurations and anomalous phases}

\begin{figure}[h!]
\centering
\includegraphics{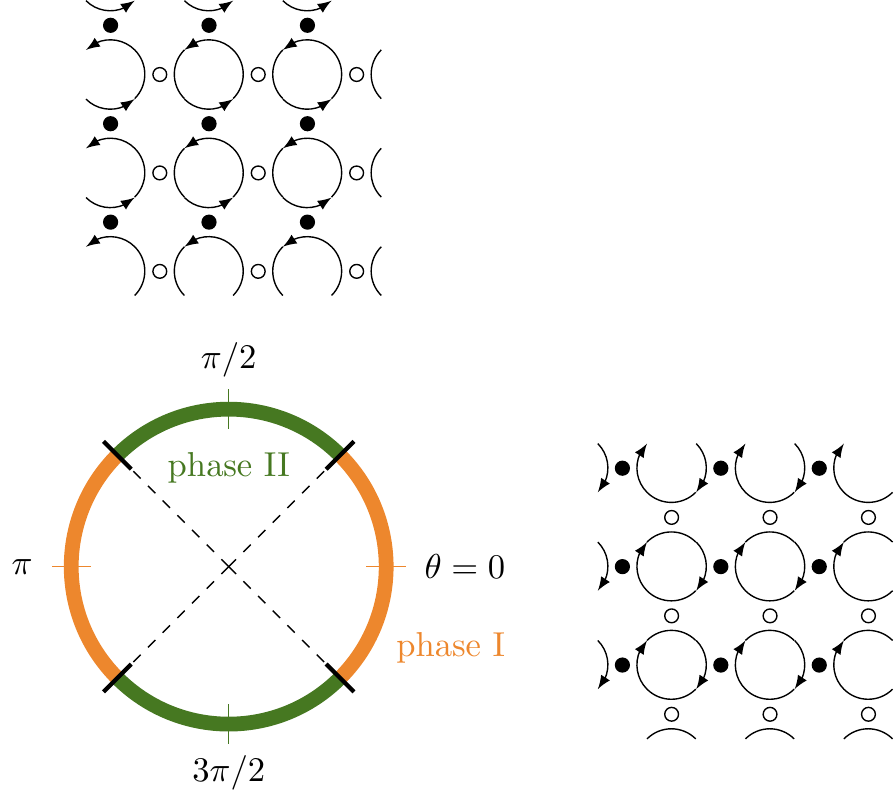}
\caption{\strong{Phase diagram and loops configurations of the L-Lattice.} (Do not confuse with the phase \emph{spectrum} of figures \ref{Floquet_eigenvalue_rotation_invariant_spectra} and \ref{relation_S_Ufn}.) The L-lattice hosts two gapped phases with a transition at $\theta=\pi/4$. When varying $\theta$, each of these phases can be continuously deformed into lattices of clockwise ($\theta=0$) or anti-clockwise ($\theta=\pi/2$) loops, that both satisfy a phase rotation symmetry with one band in the fundamental domain.}
\label{fig:L-loops}
\end{figure}

When the scattering matrices $U_j$ correspond to full reflection or full transmission, they do not split an incoming wave packet. In this situation, they describes a \emph{classical} or \emph{ballistic} propagation (as opposed to a wave-like propagation).
In the L-lattice, such a behavior arises at two special points of the phase diagram (do not confuse with the phase spectrum), when $\theta=0$ or $\theta=\pi/2$ (see figure~\ref{fig:L-loops}). Here, we observe that the network is composed only of small loops, and the corresponding point of the phase diagram is therefore called a \strong{classical loop configuration}. Notably, such loops rotate clockwise in phase I and counter-clockwise in phase II. Away from the classical configurations, the network model can be understood as a \emph{superposition} of more complicated loop configurations, where the loops now extend over several unit cells. The direction of rotation of such loops is preserved all over the gapped phase, and the transition at $\theta=\pi/4$ between clockwise and counter-clockwise phases is marked by a percolation of the possible trajectories, which allows for a path through the entire system. 

\smallbreak

Notably, a strong version of the phase rotation symmetry is satisfied at the points at the classical loop configurations, which ensures that the band structure at those points is either trivial or anomalous, a property which extends to the entire gapped phase, as topological invariants cannot change unless a gap closes. In these two situations, a unitary operator $\rot_\theta$ can be found so that
\begin{equation}
\rot_\theta\scat(k) \rot^{-1}_\theta=\ii \scat(k)
\end{equation}
with
\begin{equation}
\rot_{0} = \begin{pmatrix}
-\sigma_z & 0 \\
0              & \ii \sigma_z
\end{pmatrix}
\qquad
\text{and}
\qquad
\rot_{\pi/2}= \begin{pmatrix}
-\sigma_z & 0 \\
0              & -\ii \sigma_z
\end{pmatrix}
\end{equation}
meaning that there is only one band in the fundamental domain of the phase rotation symmetry. As shown in section~\ref{sec:consequences}, this directly implies the vanishing of the first Chern number of each band. In the section~\ref{sec:anomalous_loop_configurations}, we will see that such classical loop configurations provide, along with phase rotation symmetry, a valuable tool to design anomalous phases in network models.

In the following, we first generalize this set of observations to a more general class of scattering networks, \emph{cyclic oriented networks} (section~\ref{sec:cyclic}). Their precise definition allows us to elucidate the correspondence between the Ho-Chalker-like description and the reduced Floquet-like description (section~\ref{sec:two_pow}), which sets the ground for a proper definition of bulk topological invariants for this class of network models (section~\ref{sec:topo}). As a byproduct, we also propose a standard way to define topological invariants for a stepwise (or \enquote{discrete time}) evolution.

\subsection{Cyclic oriented networks and the phase rotation symmetry}
\label{sec:cyclic}

\subsubsection{The structure constraint}

The orientation of the links of the L-lattice is such that a wave packet traveling on the network will encounter the nodes $U_1$ and $U_2$ in a cyclic way during its evolution, namely, in a periodic sequence of the form $\cdots \to U_2 \to U_1 \to U_2  \to U_1 \to \cdots$ (there are, for example, no $U_1 \to U_1$ in this sequence). From the point of view of the wave packet, the situation is similar to a stepwise evolution periodic in time, similar to a Floquet dynamics with a (Bloch)-Floquet operator $U_{\text{F}}=U_2(k)U_1(k)$. As we shall see, there is indeed a mapping between a particular class of network models that generalize the L-lattice and stepwise Floquet evolutions.

\bigbreak

\def\sizenode{b}

A \strong{cyclic oriented network} is a (space-periodic) oriented network where any path along the directed edges is \emph{constrained} to travel through a periodic sequence of the nodes, always in the same order $\cdots \to U_s \to U_1 \to U_{2} \to \cdots \to U_{s-1} \to U_s \to U_1 \to \cdots$, where $U_j \in U(\sizenode)$ describes the scattering events at the corresponding node. A unit cell of such a network consists in $s$ nodes and $\sizenode \times s$ oriented links (in the examples, we will always consider $\sizenode=\num{2}$).
As we shall see, such a network model can be mapped to a time-periodic stepwise evolution composed of $s$ unitary operations $U_n \in U(\sizenode)$.

\smallbreak

Let us denote by $a_n, b_n, c_n, \dots$ the incoming amplitudes at the node $U_n$, and by $a_{n+1}, b_{n+1}, c_{n+1}, \dots$ the outgoing amplitudes at the same node (which are, on the cyclic network, the incoming amplitudes on the next node $U_{n+1}$). In reciprocal space, the Ho-Chalker evolution operator of such a network then reads
\begin{equation}
\mathcal{S}(k)=
 \begin{pmatrix}
  0 & 0 & \cdots & U_s(k) \\
  U_1(k) & 0 & \cdots & 0 \\
  \vdots  & \ddots  & \ddots & \vdots  \\
 0 &  \cdots & U_{s-1}(k) & 0 
 \end{pmatrix}  \qquad \in U(\sizenode \times s) 
\label{eq:S}
\end{equation}
in the Bloch basis $(a_1(k),b_1(k), a_2(k), b_2(k), \dots a_s(k),b_s(k))$.

As for the L-lattice, the form of $\scat(k)$ in this well-chosen basis is reminiscent of the cyclic structure of the oriented network. We interpret it as stemming from the existence of a \strong{structure constraint}   
\begin{equation}
D \scat(k) D^{-1} =  \ee^{\ii 2\pi/s} \scat(k) 
\label{eq_structure_constraint}
\end{equation}
where $D$ is the block-diagonal unitary matrix that reads
\begin{equation}
D = \text{diag}(1,\ee^{\ii 2\pi/s}, \ee^{\ii 4 \pi/s}, \dots , \ee^{\ii 2(s-1)\pi/s}) \otimes \Id_{\sizenode} \in U(\sizenode \times s)
\label{lattice_constraint_operator}
\end{equation}
in the same basis as $\scat(k)$, which is the standard phase rotation operator \eqref{eq:standard_prs} that satisfies $D^s=\Id$. Although the explicit expressions~\eqref{eq:S} and~\eqref{lattice_constraint_operator} for the Ho-Chalker evolution operator and its symmetry might depend on the basis and unit cell choices, they will be modified in a covariant way so that constraint~\eqref{eq_structure_constraint} will always be preserved.

Cyclic oriented networks with a given number $s$ of non-equivalent nodes and $b$ of incoming links per node define an equivalence class of networks models (where the connectivity of the underlying graph is fixed). 
The structure constraint \eqref{eq_structure_constraint} implements the restriction to this equivalence class at the level of the Ho-Chalker evolution operators $\scat(k)\in U(b\times s)$ in Bloch representation, and evolutions that preserve equation \eqref{eq_structure_constraint} therefore stay in the corresponding class.

Indeed, the structure constraint is a particular case of the phase rotation symmetry~\eqref{eq:constraint} where $\rot=D$ and with $M=s$, and the cyclic form~\eqref{eq:S} of the evolution operator highlights the reduction in the number of degrees of freedom enabled by the existence of the phase rotation symmetry\footnote{The number of degrees of freedom is reduced from $s^2 \, {\sizenode}^2$ for a generic unitary matrix to $s \, {\sizenode}^2$ when it is taken into account.}. As a consequence, the spectrum of $\scat$ is redundant: more precisely, it is obtained by $s-1$ successive rotations of the spectrum contained in a fundamental domain of length $2 \pi/s$. Moreover, the total first Chern number of the bands of $\scat(k)$ in such a fundamental domain vanishes. This set of properties will allow us to develop a mapping between the network model and a stepwise Floquet evolution. To do so, the first step is to relate the spectrum of the Ho-Chalker evolution operator $\scat$ to the spectrum of an associated Floquet evolution operator.

\subsubsection{Two points of view: simultaneous steps and sequence of steps}
\label{sec:two_pow}

\begin{figure}[htb]
  \centering
 \includegraphics{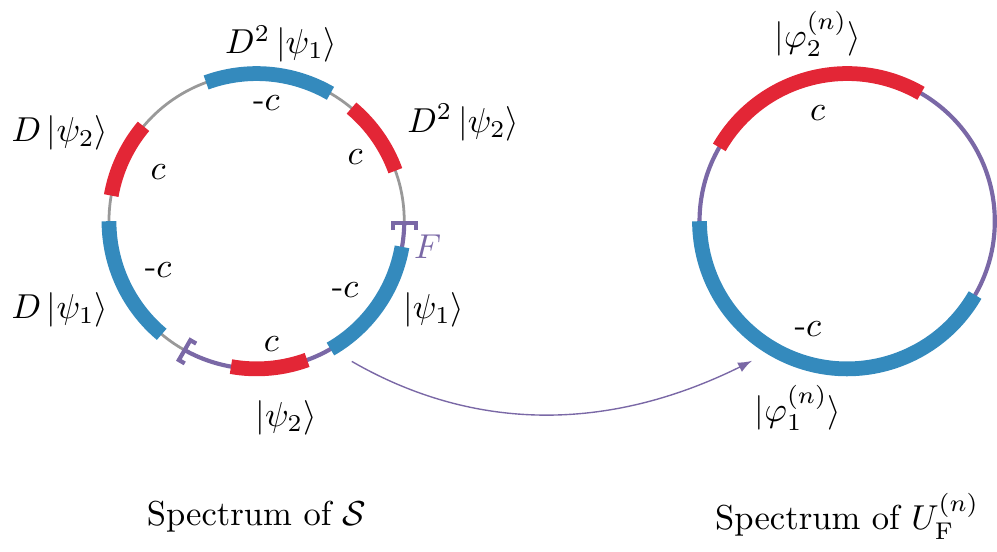}
   \caption{\label{relation_S_Ufn}\strong{Relation between the spectra of $\scat$ and $\UF^{(n)}$.} The spectrum of $\scat$ restricted to a fundamental domain $F$ of the phase rotation constraint corresponding to the structure constraint is in direct correspondence with the (full) spectra of all blocks $\UF^{(n)}$ of the repeated evolution operator $\scat^s$. By phase rotation symmetry $D$, the first Chern number on the fundamental domain $F$ is zero, and thus the two bands of $F$ ave opposite first Chern numbers.}
  \label{fig:correspondence}
\end{figure}

The particular form~\eqref{eq:S} of the Ho-Chalker evolution operator $\scat$ imposed by the structure constraint~\eqref{eq_structure_constraint} implies that its $s$-th power $\scat^s$ is block-diagonal and reads
\begin{equation}
\label{scat_power_s_block_diagonal}
\scat^s = \text{diag} ( U_{\text{F}}^{(1)}, U_{\text{F}}^{(2)}, \cdots, U_{\text{F}}^{(s)} ) 
\end{equation}
in the same basis as equation~\eqref{eq:S}, where $U_{\text{F}}^{(n)} \in U(\sizenode)$ denotes the cyclic permutation of the Floquet operator starting at step $n$, namely
\begin{equation}\label{def_UFn}
U_{\text{F}}^{(n)} = U_{n-1} \cdots U_2 U_1 U_s  \cdots U_{n+1} U_n \, .
\end{equation}
The restriction to a fundamental domain of the spectrum of the Ho-Chalker operator $\scat$ is identical to the spectrum of the Floquet operators $\UF^{(n)} \in U(\sizenode)$, up to a constant scaling factor, as illustrated in figure~\ref{relation_S_Ufn}. In this sense, $\scat$ can be \emph{reduced} to the smaller-dimensional operator $\UF^{(n)}$. The eigenstates of the $\UF^{(n)}$ can be obtained from the eigenstates of $\scat$. The converse is not fully possible without the knowledge of the matrices $U_j(k)$, but we will see that the first Chern numbers of the bands of any of the $\UF^{(n)}$ (for any given $n$) entirely determine the ones of $\scat$.

\smallbreak

Let $\ket{\psi}$ be an eigenstate of $\scat$ with eigenvalue $\lambda$, so that $\scat \ket{\psi} = \lambda \ket{\psi}$, and thus $\scat^s \ket{\psi} = \lambda^s \ket{\psi}$. Decomposing the vector $\ket{\psi}$ into $s$ smaller vectors $\ket{\varphi^{(r)}}$ 
as
\begin{equation}
  \ket{\psi} = \left(  \ket{\varphi^{(1)}}, \cdots , \ket{\varphi^{(s)}} \right)^T
  \label{correspondence_varphi_psi}
\end{equation}
it follows from~\eqref{eq:S} that
\begin{equation}
U_n \ket{\varphi^{(n)}} = \lambda \ket{\varphi^{(n-1)}} 
\label{eq:recursive_phi}
\end{equation}
and we infer from equation~\eqref{scat_power_s_block_diagonal} the eigenvalue equation for the Floquet operators
\begin{equation}
  U_{\text{F}}^{(n)} \ket{\varphi^{(n)}} = \lambda^s \ket{\varphi^{(n)}} .
\end{equation}
Importantly, the phase spectrum of $U_{\text{F}}^{(n)}$ does not depend on $n$, meaning that the Floquet spectrum is invariant under a change of the origin of time, as expected. This construction can be applied to the set of $\sizenode \times s/s=\sizenode$ eigenvectors $\ket{\psi_j}$ of $\scat$ with eigenvalues $\lambda_j$ in the fundamental domain $F$ to obtain two linearly independent eigenstates $\ket{\varphi_j^{(n)}}$ of $U_{\text{F}}^{(n)}$. As a consequence, we have on the one hand
\begin{equation}\label{eq:spec_scat}
  \scat = \sum_{r=0}^{s-1} \sum_{j=1}^{\sizenode} \ee^{-\ii 2 \pi r/s} \lambda_j  D^{r} \ket{\psi_j}\!\bra{\psi_j} D^{-r}
\end{equation}
and on the other hand
\begin{equation}
  U_{\text{F}}^{(n)} = \sum_{j=1}^{\sizenode}  \lambda_j^s  \ket{\varphi_j^{(n)}}\!\bra{\varphi_j^{(n)}}
  \label{eq:spec_UF}
\end{equation}
where the correspondence between $\ket{\psi_j}$ and $\ket{\varphi_j^{(n)}}$ is given by~\eqref{correspondence_varphi_psi} and illustrated in figure~\ref{fig:correspondence}.

\smallbreak

Indeed, the \emph{complete} correspondence between the Ho-Chalker description and the Floquet description involves, on one side, the Ho-Chalker evolution operator $\scat(k)$ and, on the other side, the stepwise Floquet \emph{evolution} with steps $(U_1, \dots, U_{s})$ [as opposed to only the Floquet operator $U_{\text{F}}^{(n)}$, from which it is not possible to reconstruct $\scat(k)$ entirely]. In particular, both points of view allow for a complete topological characterization of the system. However, we have seen that the phase spectrum of the Floquet operator $U_{\text{F}}^{(n)}$ is enough to reconstruct the phase spectrum of $\scat$, and we will see in the next paragraph that this is also true for the first Chern numbers of their bands.

\subsubsection{Consequences on the first Chern numbers}
\label{sec_consequences_first_chern_number_networks}

We have seen that the spectra of the Ho-Chalker operator $\scat(k)$ and of the Floquet operator $U_{\text{F}}^{(n)}$ are in direct correspondence, and can be obtained from one another, possibly up to a constant phase. In addition, the first Chern numbers of their bands are also in direct correspondence. More precisely, let $P^{\text{F}}_{\eta, \eta'}$ be the projector on states between the gaps $\eta$ and $\eta'$ of $U_{\text{F}}^{(n)}$, and let $P^{\text{HC}}_{\eta/s, \eta'/s}$ be the projector on states between the gaps $\eta/s$ and $\eta'/s$ of $\scat$. Then, 
\begin{equation}
  \label{eq:chern_HC_Floquet}
  \cc(P^{\text{F}}_{\eta, \eta'}) = \cc(P^{\text{HC}}_{\eta/s, \eta'/s}).
\end{equation}
Although a more direct proof could be devised, we infer this identity from our results on the complete topological characterization of network models which is discussed in the last section, and in particular from equation~\eqref{equality_W} (which is proven in appendix~\ref{app:proofW}) and the relation~\eqref{relation_W_chern}.

\bigbreak

A particular but typical situation arises when all bands are well-defined and composed of only one state. Then, the spectrum of $\scat$ is composed of $\sizenode \times s$ bands separated from each other by $\sizenode \times s$ gaps. Due to the phase rotation symmetry, it is sufficient to consider the $\sizenode$ bands in a fundamental domain described by projectors $P[\psi_j] = \ket{\psi_j}\bra{\psi_j}$, with $j = 1, \dots, \sizenode$. The Floquet operator $U_{\text{F}}^{(n)}$ has also $\sizenode$ bands corresponding to projectors $P[\varphi_j^{(n)}] = \ket{\varphi_j^{(n)}}\bra{\varphi_j^{(n)}}$, and equation~\eqref{eq:chern_HC_Floquet} simplifies into
\begin{equation}
\cc\left(P[\psi_j]\right) = \cc(P[\varphi_j^{(n)}]).
\label{eq:chern_HC_Floquet2}
\end{equation}
This illustrates that the first Chern number of a band $j$ of a generalized Ho-Chalker operator is simply obtained from any of its associated Floquet operators $U_{\text{F}}^{(n)}(k)$, as sketched in figure~\ref{fig:correspondence}. Equation~\eqref{eq:chern_HC_Floquet2} is of practical importance, since $U_{\text{F}}^{(n)}$ has a smaller dimension than that of $\scat$.

\smallbreak

To obtain a vanishing first Chern number phase (where $\cc(P[\psi_j])=0$ for all of the $\sizenode \times s$ bands of $\scat$), it is therefore enough to show that the Floquet operator $U_{\text{F}}^{(n)}$ has a vanishing first Chern number phase (where $\cc(P[\varphi_j^{(n)}])=0$). This is far easier, as we have to deal with $U(\sizenode)$ matrices instead of larger $U(\sizenode \times s)$ matrices. As we have seen in section~\ref{sec:consequences}, this is achieved when $U_{\text{F}}^{(n)}$ is endowed with a (strong) phase rotation symmetry~\eqref{eq:constraint}, with only one band in the fundamental domain, that is to say, when there is a unitary operator $\rot \in U(\sizenode)$ such that 
\begin{equation}
\rot U_{\text{F}}^{(n)}(k) \rot^{-1} = \ee^{\ii 2\pi /b} U_{\text{F}}^{(n)}(k) \ .
\label{eq:chern0}
\end{equation}
For $b=2$, such a constraint is similar to the \enquote{phase shift} property pointed out by Asb{\'{o}}th and Edge in two-dimensional discrete-time quantum walks~\cite{Asboth2015}. 

\section{Vanishing first Chern number phases and classical loop configurations}
\label{sec:anomalous_loop_configurations}

\subsection{A procedure to identify vanishing first Chern number phases in network models}

The Ho-Chalker evolution operators $\scat$ of cyclic oriented networks always have a phase rotation symmetry with $\rot=D$, so that there are $\sizenode$ bands in the fundamental domain (in this section, we will always consider situations where $\sizenode = 2$). This allows one to reduce the dimension of the problem and map it onto a Floquet dynamics. However, this does not guarantee the vanishing of the first Chern numbers. To obtain anomalous phases where all first Chern numbers vanish, an extra condition has to be found on the Floquet operator, such as equation~\eqref{eq:chern0}, where another phase rotation symmetry applies to $U_{\text{F}}^{(n)}$.
This approach is tantamount to the one consisting in directly finding out a \enquote{stronger} phase rotation symmetry for the Ho-Chalker evolution operator $\scat$ with only \emph{one} band in the fundamental domain, as discussed in section~\ref{sec:L_lattice} on an example. Yet, it is usually convenient to work in the Floquet point of view where smaller matrices are involved, as discussed in section~\ref{sec_consequences_first_chern_number_networks}. 

\bigbreak

In this section, we introduce a simple qualitative method to establish whether a cyclic oriented network has a vanishing first Chern number phase or not.
Our analysis lies on two points. 

First, we identify the possible classical loops configurations, as we did for the L-lattice (see section~\ref{sec:L_lattice}). These configurations are obtained by considering the possible loops in the unit cell when the nodes are either fully transmitting or fully reflecting. Intuitively, these configurations ensure that the phase being described is gapped, since the amplitude of a state cannot escape from a loop to propagate in the network.

Second, we associate a Floquet operator $U_{\text{F}}^{(n)}$ to each classical loop configuration (as the the first Chern numbers do not depend on the choice of the starting node, we can arbitrarily choose one of them). Importantly, the Floquet operator is a product of either diagonal or anti-diagonal step operators $U_n$, because of the classical loop structure, and it is therefore itself either diagonal or anti-diagonal. 
Depending on its form, one can possibly conclude about the existence of a phase rotation symmetry~\eqref{eq:chern0} for the Floquet operator by easily exhibiting a suitable phase rotation operator $\rot$.
In particular, if $U_{\text{F}}$ is anti-diagonal, then equation~\eqref{eq:chern0} is always satisfied with $\rot=\sigma_z$, which guaranties the vanishing of the first Chern number of bands of the Floquet operator, and therefore the vanishing of the first Chern number of the bands of the Ho-Chalker evolution operator. 

Let us now apply this analysis to concrete cyclic oriented networks. 

\subsection{The L-lattice ($s=2$)}

The two loops configurations of the L-lattice have already been discussed in section~\ref{sec:L_lattice} (see figure~\ref{fig:L-loops}), where we exhibited a rotation phase symmetric operator for the Ho-Chalker evolution operator. As discussed above, one can equivalently consider any of the associated Floquet operator $U_{\text{F}}^{(n)}$. In phase I, a loop corresponds to the sequence $a_1 \rightarrow b_2 \rightarrow b_1, \rightarrow a_2 \rightarrow a_1$ meaning that $U_1$ is anti-diagonal (it changes $a$ to $b$ and $b$ to $a$) and $U_2$ is diagonal. Their product is thus anti-diagonal, and the first Chern numbers therefore vanish. In phase II, a loop corresponds to the sequence $a_1 \rightarrow a_2 \rightarrow b_1 \rightarrow b_2 \rightarrow a_1$, meaning that $U_1$ is diagonal and $U_2$ is anti-diagonal. Again, their product is anti-diagonal and the first Chern numbers vanish. This is of course in agreement with the analysis of the Ho-Chalker operator done in section~\ref{sec:L_lattice}. 

This reasoning can now be applied to cyclic oriented networks beyond the L-lattice. 

\subsection{The oriented Kagome lattice ($s=3$)}

\begin{figure}[!h]
\centering
\includegraphics{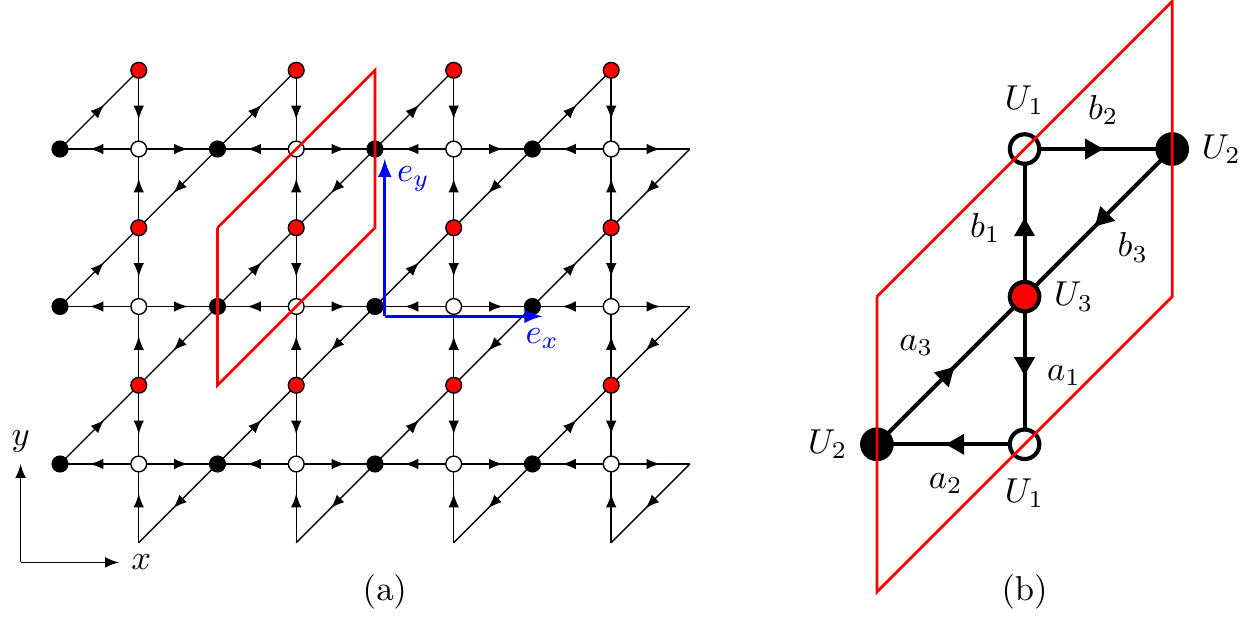}
\caption{\strong{Oriented Kagome lattice.} (a) Oriented Kagome lattice with $6$ inequivalent links and $3$ inequivalent nodes per unit cell enhanced in red and detailed in (b).}
\label{fig:kagome}
\end{figure}

The cyclic oriented network with $s=3$ corresponds to a Kagome lattice shown in figure~\ref{fig:kagome} (a). In this case, the unit cell is composed of $s=3$ inequivalent nodes $U_j$ ($j \in [1,s]$) and $2s=6$ inequivalent oriented links denoted by $(a_j, b_j)$ [see figure~\ref{fig:kagome} (b)]. Note that the oriented Kagome lattice has been considered to describe arrays of optical coupled resonators arranged in a honeycomb lattice by Pasek and Chong~\cite{Pasek2014}. 

This oriented network allows for different possible loops configurations. 
Let us identify some of those which necessarily correspond to a vanishing first Chern number phase. Following the method discussed above, we select loops such that the product of the three $U_j$'s is anti-diagonal. As previously, $U_j$ is anti-diagonal if it changes $a\leftrightarrow b$ and is diagonal otherwise.
\begin{figure}[!t]
\centering
\includegraphics{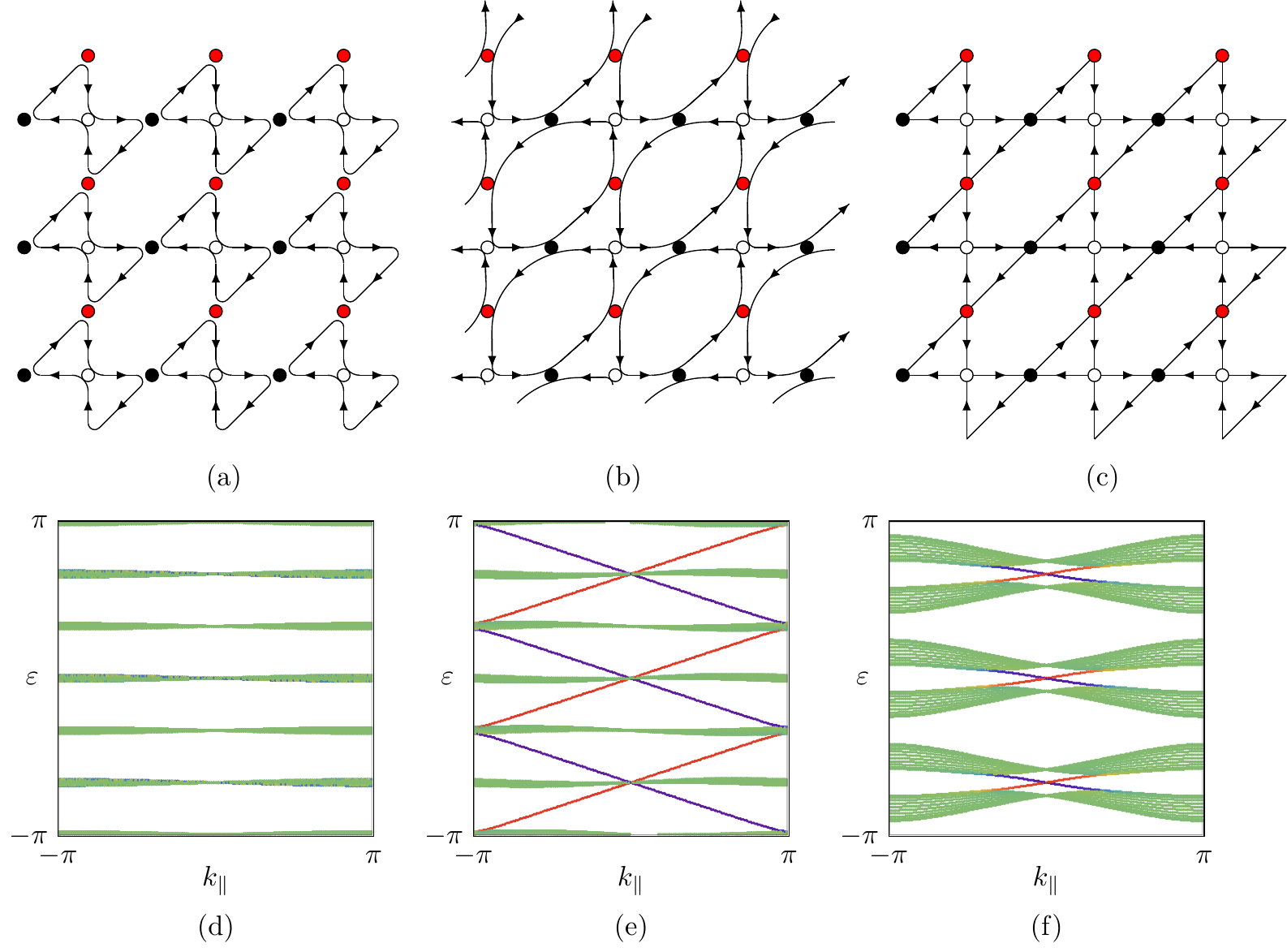}
\caption{\strong{Examples of loops configurations in the oriented Kagome lattice} (a) and (b) Loops configurations that display a vanishing first Chern number phase. The phase spectra (d) and (e) for these configurations in a ribbon geometry confirm this result, and reveal that (a) corresponds to a trivial topological phase (with no edge state in the gaps) whereas (b) corresponds to an anomalous topological one (with one chiral edge state per edge in each gap). The configuration (c) displays no loop so that only the first Chern number on a fundamental domain is constrained to vanish. The phase spectrum on a strip geometry (f) confirms this result.}
\label{fig:kagome_loops}
\end{figure}
With this in mind, it is clear that the two configurations represented in figures ~\ref{fig:kagome_loops} (a) and (b) correspond to a vanishing first Chern number phase.
Indeed, for the loops sketched in figure ~\ref{fig:kagome_loops} (a),  $U_1$ is anti-diagonal whereas $U_2$ and $U_3$ are diagonal, and for the loops shown in figure~\ref{fig:kagome_loops} (b),  all the $U_j$'s are anti-diagonal.
These results are confirmed by a direct diagonalization of the phase spectrum in a strip geometry which exhibits, for each of these two configurations, an equal (algebraic) number of edge states in each gap [$0$ in the spectrum (d) and $1$ per edge in the spectrum (e) of the figure~\ref{fig:kagome_loops}], as expected for a vanishing first Chern number phase.
In contrast, if one considers a case where any incoming amplitude to a node is partially scattered  onto each outgoing link, then this does not correspond to a loops configuration and $U_{\text{F}}$ is neither diagonal nor diagonal [see figure~\ref{fig:kagome_loops} (c)]. Thus, provided such a phase is gapped, the first Chern numbers may \emph{not vanish}, as shown in figure~\ref{fig:kagome_loops} (f).

This analysis gives one an insight on the control of the first Chern number. However, to discriminate a topologically trivial phase [figure~\ref{fig:kagome_loops} (d)] from an anomalous topological one (with zero first Chern number and edge states [figure~\ref{fig:kagome_loops} (e)]), it is still required to compute the topological invariants defined in section~\ref{sec:topological_invariants_network}.

\section{Topological characterization of cyclic scattering network models}
\label{sec:topo}

We now want to fully characterize cyclic scattering network models, and in particular to account for anomalous phases. Both the Ho-Chalker and the Floquet points of view provide ways to define proper bulk topological invariants, which crucially depend on the existence of the structure constraint~\eqref{eq_structure_constraint}. In both cases, we \emph{interpolate} the discrete evolution to a continuous one, in order to use the tools of homotopy theory. A first step in this direction is to properly define topological invariants for a stepwise evolution; the section~\ref{sec:topological_invariants_swe} is devoted to this task. Equipped with this tool, it is possible to actually define topological invariants for the network model in section~\ref{sec:topological_invariants_network}. In the Ho-Chalker point of view, we require the interpolation to always satisfy the structure constraint of the oriented network to be able to define meaningful invariants. No such requirement is necessary in the Floquet point of view, as the reduction to the stepwise dynamics already takes the structure constraint into account. Indeed, both points of view are equivalent, and the corresponding invariants can be related one to another.

\subsection{Topological characterization of a stepwise evolution}
\label{sec:topological_invariants_swe}

A topological characterization of periodically driven systems was proposed by Rudner, Lindner, Berg, and Levin~\cite{Rudner2013} for systems without specific symmetries in two dimensions. A topological invariant $W_{\eta}$ can be assigned to each spectral gap $\eta$ of the Bloch-Floquet operator $U(t=T,k)$, thus remarkably establishing a new bulk-boundary correspondence for periodically driven systems~\cite{Rudner2013}. This index $W_{\eta}[U]$ is defined as the degree (or winding number) of a \enquote{periodized evolution operator} $V_{\eta}(t,k)$ built from the full evolution operator $U(t,k)$. 
This method is applicable both to Floquet systems actually periodically driven in time, and to lattices of evanescently coupled light waveguides when the paraxial direction of propagation plays somehow the role of time~\cite{Rechtsman2013}. In contrast, stepwise evolutions like~\eqref{eq:Usequence} are not continuous maps, as opposed to the usual evolution operator~\eqref{eq:def_U}. Hence, the index $W$ is not directly applicable to such evolutions. 

In the following, we propose a systematic procedure to extend Rudner \emph{et al.}'s~\cite{Rudner2013} $W$ index to discrete evolutions. In order to do so, an interpolating continuous-time evolution must be associated to any stepwise evolution $U = U_{N} \cdots U_{1}$. The main idea is that when the stepwise evolution is correctly specified, one can assume that each step $U_{n}$ is generated by a time-independent Hamiltonian $H_{n}$, so that $U_{n} = \ee^{-\ii \tau_n H_{n}}$ for some duration $\tau_n$. The description of the stepwise evolution does not contain more information.
A similar method, where a Hamiltonian realizing the discrete-time quantum walk is explicitly constructed, was described by Asbóth and Edge~\cite{Asboth2015}. 

\subsubsection{Topological characterization of a continuous unitary evolution}

In this paragraph, we review the construction by Rudner et al.~\cite{Rudner2013} of a topological invariant $W_{\eta}$ for unitary evolutions, and define all the tools required for the upcoming construction.

We start with a continuous unitary evolution $U(t,k)$ from an origin time $t=0$ to a finite time $t=T$, and assume that $U(T)$ is gapped. It is then convenient to define a time-independent \emph{effective Hamiltonian} $H^{\text{eff}}(k)$. In the context of Floquet theory of periodically driven systems, such an effective Hamiltonian would generate the \enquote{stroboscopic evolution} at discrete times $U(n T) = [U(T)]^{n}$. Namely, we want that $U(T,k) = \ee^{-\ii \, T H^{\text{eff}}(k)}$. A crucial point of reference~\cite{Rudner2013} is that the effective Hamiltonian is not unique, as it is defined as a logarithm of the Floquet operator. More precisely, the branch cut $\eta$ of the logarithm must be chosen in a spectral gap of the Floquet operator $U(T)$, to define (e.g., by spectral decomposition)
\begin{equation}\label{def:Heff}
	H^{\text{eff}}_{\eta}(k) = \frac{\ii}{T} \log_{-\eta} U(T,k)
\end{equation}
where the complex logarithm with branch cut along an ray with angle~$-\eta \in \RR$ is defined as
\begin{equation}
\label{eq_def_log}
	\log_{-\eta}(\ee^{\ii \varphi}) = \ii \varphi
	\quad
	\text{for}
	\quad
	- \eta - 2 \pi < \varphi < - \eta.
\end{equation}
The periodized evolution operator is then defined as\footnote{Although different from the original one from \cite{Rudner2013}, this definition is equivalent and leads to the same topological invariant, see \cite[Appendix C]{Carpentier2015}.} 
\begin{equation}
V_{\eta}(t,k) = U(k,t) \ee^{\ii t H^{\text{eff}}_{\eta}(k)}.
\label{eq:def_V}
\end{equation}
Finally, as $V_{\eta}$ is periodic both in time and on the Brillouin zone $\BZ$, the bulk topological index is defined as its degree or winding number 
\begin{equation}\label{def_W}
W_{\eta}[U] \equiv \deg(V_\eta) \in \ZZ,
\end{equation}
where the degree of a periodic map is formally defined as
\begin{equation}\label{degreeformula}
\deg(V_\eta) \equiv \frac{1}{24\pi^2} \int_{[0,T]\times \BZ} \left(V_\eta \right)^* \chi
\qquad
\text{where}
\qquad
\left(V_\eta \right)^* \chi = \tr\left[ (V_\eta^{-1} \dd V_\eta)^3\right].
\end{equation}

When $\eta$ and $\eta'$ are in the same spectral gap of $U(T)$ (called the quasi-energy spectrum in the context of periodically driven systems), then $W_{\eta}[U] = W_{\eta'}[U]$ (and indeed, $W_{\eta+2\pi}[U]=W_{\eta}[U]$). When there are several gaps in the phase spectrum, however, there are as many topological invariants defined for the unitary evolution.

Remarkably, the interface between two driven systems with bulk evolution operators $U_{\text{left}}$ and $U_{\text{right}}$ carries $n_{\text{es}}(\eta)$ topologically protected chiral edge states (counted algebraically with chirality) in the gap of quasi-energy $\eta$ with \cite{Rudner2013} 
\begin{equation}
	n_{\text{es}}(\eta) = W_{\eta}[U_{\text{left}}] - W_{\eta}[U_{\text{right}}].
\end{equation}

At equilibrium, the first Chern numbers of energy bands are sufficient to characterize the topology of quantum Hall like systems. In a periodically driven system, the quasi-energy bands can also carry a nonzero first Chern number. Rudner et al.~\cite{Rudner2013} showed that even though the data of all the first Chern numbers is not sufficient to fully characterize the topology of Floquet states, they are still significant, and give the variation in the $W$ invariant between the gaps above and below the band. More precisely, let $- 2\pi < \eta_1, \eta_2 < 0$ be two quasi-energies and $P_{\eta_1,\eta_2}(k)$ the spectral projector on states with quasi-energy between $\eta_1$ and $\eta_2$, i.e. on eigenstates with eigenvalues $\ee^{-\ii \eta}$ in the arc joining $\ee^{-\ii  \eta_1}$ and $\ee^{-\ii  \eta_2}$ clockwise on the circle $U(1)$. The difference between the gap invariants $W$ is then related to the first Chern number $\cc$ of the quasi-energy band in between by
\begin{equation}
	\label{relation_W_chern}
	W_{\eta_2}[U] - W_{\eta_1}[U] = - C_{1}(P_{\eta_1,\eta_2}).
\end{equation}

\subsubsection{Interpolating a discrete-time evolution}
\label{sec:interpolating_swe}

In a stepwise evolution, one only knows the evolution operator at several discrete times. We would like to interpolate such data to a physically relevant continuous evolution. As such an interpolation is not unique, we need a specification to construct it in a unique way, or at least to get equivalent interpolations from the point of view of topological properties. 
The main idea is that the choice of the interpolation should not \enquote{add} or \enquote{remove} anything to the topology.

Roughly speaking, each step $U_i$ should be interpolated from the identity $\Id$ through an evolution of the form $\mathcal U_{\text{int}} = \ee^{-\ii t H_i}$ for some effective Hamiltonian $H_i$. The choice of an interpolation where the phases grow linearly is, however, a natural choice in this context~\cite{Nathan2015}, as it necessarily corresponds to a trivial evolution when all first Chern numbers of $H_i$ vanish. In this situation all such interpolations are actually equivalent. In the general case where the step operators may carry nonzero first Chern numbers, we must assume that each step operator stems from a constant Hamiltonian, and moreover that the evolution was sufficiently short with respect to the characteristic time scales of the Hamiltonian. This hypothesis is necessary to accurately interpolate the discrete-time evolution, as it ensures that the gap $\eta = -\pi$ of the step operator is trivial. Without this additional information, there are not enough data to unambiguously reconstruct the evolution (essentially because it is not sufficiently discretized to capture all the physical information of the system). 

\bigbreak

We first consider the evolution operator generated by a (known) constant Hamiltonian, i.e. to do the one-step version in a situation where the result is known. A time-independent Hamiltonian $H$ generates a step evolution operator $U_{\text{F}} = U_1 = \ee^{-\ii T H}$. In this case, we already know that the correct interpolation is indeed
\begin{equation}
	U(t) = \ee^{-\ii t H}.
\end{equation}
To express it in terms of the step operator $U_1$ only, we use the effective Hamiltonian, defined in~\eqref{def:Heff}, 
corresponding to $U_1$ and with logarithm branch cut $\eta = -\pi$. We have
\begin{equation}
	H^{\text{eff}}_{\eta=-\pi}[\ee^{-\ii H T}] = H
\end{equation}
at least for a small enough\footnote{To be precise, the maximum energy in absolute value $\max | \sigma(H) |$ of $H$ should be smaller than $\pi/T$, or $h/2T$ if we restore the Planck constant. In this way, there is always a gap around the eigenvalue~$\ee^{-\ii \pi}$.} $T$. We may then interpolate the evolution ending with the step operator $U_1$ by
\begin{equation}\label{eq:def_standardHeff}
	\mathcal{U}_{\text{int}}[U_1](t) \equiv \exp\left( -\ii t H^{\text{eff}}_{-\pi}[U_1] \right).
\end{equation}
and we see that this formula immediately generalizes to any step operator $U_1$, even without knowing some underlying time dynamics. Besides this definition being natural as the choice of cut, $\eta = -\pi$ ensures that when $U_1 = \ee^{-\ii T H}$ (for $T$ small enough) we recover
\begin{equation}
	\mathcal{U}_{\text{int}}[U_1](t) = \ee^{-\ii t H}.
\end{equation}
This is obviously true in $t=T$ whatever the choice of $\eta$ is, but it is not necessarily valid for intermediate times. Besides with this choice,
\begin{equation}\label{eq:W_Heff_C1}
	W_{\eta}[\mathcal{U}_{\text{int}}[U_1]] = \deg[\exp\left( -\ii t H^{\text{eff}}_{-\pi}[U_1] \right) \cdot \exp\left( \ii t H^{\text{eff}}_{\eta}[U_1] \right)] = -\cc(P_{-\pi,\eta})
\end{equation}
correctly accounts for the topology of the time-independent system (e.g. for $\eta=0$ it gives, up to the usual sign, the first Chern number of the valence band). 

\bigbreak

\begin{figure}[htb]
  \centering
\includegraphics{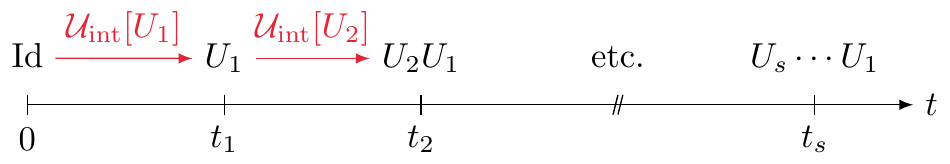}
  \caption{\label{interpolation}\strong{Construction of the interpolation of a stepwise evolution.} The evolution is constituted of $s$ time-ordered step operators $U_j$. A time $t_j$ is attributed to each step and a continuous map is systematically built to interpolate between two successive steps. It follows a continuous map $t\rightarrow U(t)$ from $\Id$ to $\UF^{(1)}$.}
\end{figure}

We now move on to the general case. When there are several steps in the stepwise evolution, a natural interpolation of the full evolution consists in concatenating the one-step interpolations (see figure \ref{interpolation}).
Thus for each step operator $U_i$ involved in a stepwise evolution, an explicit interpolation is given by definition~\eqref{eq:def_standardHeff}, as long as this operator is gapped around $-\pi$, so that it coincides with time evolution when $U_j$ actually comes from a Hamiltonian dynamics. When $U_j$ is gapped, but not at phase $-\pi$, a constant ($k$ independent) rotation of the phase spectrum can be factored out of the step operator. When the step operator has vanishing first Chern numbers, all interpolations by constant effective Hamiltonians are topologically equivalent, so any choice of such a rotation is acceptable. On the other hand, when the step operator has nonvanishing first Chern numbers, we must provide additional data for the interpolation to be unique, and the hypothesis of a trivial gap around $-\pi$ is a natural and sufficient choice. The critical case where the step operator $U_j$ is gapless is discussed in Section~\ref{subsec:gapless}.

To be precise, let us consider a time-periodic stepwise evolution of period $T$, namely, the data of several times $t_1, \dots,t_s$ and $t_s = T$ and corresponding unitary operators $U_1,\dots,U_s$. We assume that such operators are gapped at phase $-\pi$.
The evolution operator is only defined at discrete times $p T + t_j$ for $1 \leq j < s$ and $p \in \NN$ as
\begin{equation}
	U(p T + t_j) = U_{j} U_{j-1} \cdots U_1 \; (U_{\text{F}}^{(1)})^{p}
  \qquad
  \text{and}
  \qquad
  U(0) = \Id
\end{equation}
where
\begin{equation}
	U_{\text{F}}^{(1)} = U_{s} U_{s-1} \cdots U_1.
\end{equation}
Indeed, $U_{\text{F}}^{(1)}=U(T)$.
We interpolate this evolution as
\begin{equation}
\label{eq:U_interpol}
  U(t) = \mathcal{U}_{\text{int}}[U_j]\left( \displaystyle\frac{t-t_{j-1}}{t_j-t_{j-1}} \; T \right) U_{j-1} \cdots U_1 \qquad \text{for $t_{j-1} \leq t \leq t_j$}
\end{equation}
where by convention $t_0 = 0$ and $t_{s}=T$ (indeed, the first step is simply the correctly rescaled $\mathcal{U}_{\text{int}}[U_1]$).
We can then extend the previous notation by setting
\begin{equation}
	\mathcal{U}_{\text{int}}[U_1, \dots, U_s] = U(t),
\end{equation}
where it is implied that $U_1, \dots, U_s$ are in the right order. Note that the choice of times $t_j$ for the interpolation is completely arbitrary and will not influence its topological properties.

Equivalently, we may consider the piecewise constant Hamiltonian
\begin{equation}\label{eq:SWE_Heffs}
	H(t) = \mathcal{H}^{\text{eff}}_{-\pi}[U_j] \quad \text{for $t_{j-1} \leq t \leq t_j$}
\end{equation}
with $t_0=0$ and $t_s=T$, and consider the corresponding evolution operator.

\subsubsection{Topological invariants for discrete evolutions}

The previous construction allows us to \emph{define} the topological invariant associated with the stepwise evolution as the topological invariant associated with this continuous-time evolution as
\begin{equation}\label{eq:def_W_SWE}
	W^{\text{SWE}}_{\eta}[U_{\text{F}}^{(1)}] = W_{\eta}\big[ \mathcal{U}_{\text{int}}[U_1, \dots, U_s] \big].
\end{equation}
Note that functionally, $W^{\text{SWE}}_{\eta}[U_{\text{F}}^{(1)}]$ depends on the entire sequence of steps $(U_1,...,U_s)$ and not only on their product $U_{\text{F}}^{(1)}$.
This invariant is indeed associated with an \emph{ordered} sequence of steps, but is invariant under any circular permutation of the step sequence. Such permutations correspond to the Floquet operators $U_{\text{F}}^{(n)} = U_{n-1} \cdots U_2 U_1 U_s  \cdots U_{n+1} U_n$ defined in~\eqref{def_UFn}, and one has
\begin{equation}\label{eq:equal_SWE}
W^{\text{SWE}}_{\eta}[U_{\text{F}}^{(1)}] = W^{\text{SWE}}_{\eta}[U_{\text{F}}^{(n)}] \qquad \forall \, n \in {1,\ldots,s} \, .
\end{equation}
This property simply means that the choice of origin of time (i.e. the first step) is not relevant for a periodic system, similarly to Floquet systems with a continuous time evolution. We, however, give an independent and explicit proof of this property in appendix~\ref{app:proofWSWE}, based on the fact that $W^{\text{SWE}}_{\eta}$ is a degree computation [see equation~\eqref{def_W}] and hence invariant under homotopy (i.e. under smooth deformations). We show that the two periodized interpolations are homotopic, so their degrees coincide. 

\smallbreak

Although it was devised with oriented scattering networks in mind, this construction is applicable to any stepwise evolutions like discrete-time quantum walks.

\subsection{Topological invariants for cyclic network models}
\label{sec:topological_invariants_network}

Equipped with tools to define topological invariants for stepwise evolutions, we can now move on the case of cyclic scattering networks. Physically, the effective Floquet stepwise evolution essentially describes the evolution from the point of view of one wave packet traveling on the network, whereas the Ho-Chalker point of view consists in studying the global evolution of the entire network.

\subsubsection{The Floquet point of view}

As we have seen in section~\ref{sec:two_pow}, a cyclic oriented network model can be mapped to a time-periodic stepwise evolution with steps $(U_1, U_2, \dots, U_{s})$, where $k \mapsto U_j(k) \in U(\sizenode)$ are maps from the Brillouin zone to the unitary group. In this \emph{Floquet point of view}, such step operators are multiplied (in the right order) to obtain the Floquet operator $U_{\text{F}}^{(n)} = U_{n-1} \cdots U_2 U_1 U_s  \cdots U_{n+1} U_n$ associated to the network model. Hence we simply apply the results of section~\ref{sec:topological_invariants_swe} to define as a topological invariant of the network the quantity
\begin{equation}
  W^{\text{SWE}}_{\eta}[U_{\text{F}}^{(1)}] \in \ZZ
\end{equation}
defined in equation \ref{eq:def_W_SWE} which, as we said before, does actually not depend on the choice of the first step.

\subsubsection{The Ho-Chalker point of view}

We would like now to apply the same reasoning for the Ho-Chalker evolution operator $\scat(k) \in U(\sizenode \times s)$.
It is still possible to interpolate from the identity to $\scat(k)$. However, this method does not take into account the nature of the scattering network, which is expressed by the structure constraint~\eqref{eq_structure_constraint}, and is likely to fail. A strong evidence in this direction is that we do not expect to observe anomalous topological phases (with vanishing first Chern numbers) in this situation, as the interpolation effectively reproduces the effect of a constant Hamiltonian $\mathcal{H}^{\text{eff}}_{-\pi}[\mathcal{S}]$, and equation~\eqref{eq:W_Heff_C1} tells that the degree of a single-step evolution only captures first Chern numbers. As it is clear from the examples that such phases do exist (see section~\ref{sec:L_lattice}), this construction fails to capture the full topology of the network model.

To fully take into account the structure of the cyclic oriented network, we need a somehow more complex interpolation. Starting from the structure constraint~\eqref{eq_structure_constraint} for $\mathcal{S}$, which encodes the particular relations between the different links of the network, we propose instead the following interpolation:
\begin{equation}\label{def_UintD_S}
\mathcal U_{\text{int,HC}}[\mathcal{S}](t) =
 \begin{pmatrix}
  0 & 0 & \cdots & \mathcal U_{\text{int}}[U_s](t) \\
  \mathcal U_{\text{int}}[U_1](t) & 0 & \cdots & 0 \\
  \vdots  & \ddots  & \ddots & \vdots  \\
 0 &  \cdots & \mathcal U_{\text{int}}[U_{s-1}](t) & 0 
 \end{pmatrix}  \qquad \in U(\sizenode \times s).
\end{equation}
which has to be compared with~\eqref{eq:S}. Crucially, the structure constraint~\eqref{eq_structure_constraint} is satisfied all along this interpolation, namely, for every $t \in [0,T]$, we have
\begin{equation}
 D \, \mathcal U_{\text{int,HC}}[\mathcal{S}](t) \, D^{-1} =  \ee^{\ii 2\pi/s} \, \mathcal{U}_{\text{int,HC}}[\mathcal{S}](t).
\end{equation}
We then define
\begin{equation}\label{eq:def_Wonestep}
W_{\eta}^{\text{HC}}[\mathcal{S}] = W_\eta\big[\mathcal U_{\text{int,HC}}[\mathcal{S}]\big] \in \ZZ
\end{equation}
as the invariant for the $\mathcal{S}$ matrix describing the one-step evolution of a cyclic oriented network.

\subsubsection{Relation between the topological invariants: equivalence between both points of view}

So far we have defined two different topological invariants for a cyclic oriented network:
(1) $W_\eta^\text{HC}$ is associated to the one-step Ho-Chalker evolution operator $\scat$ describing the network model and 
(2) $W_\eta^{\text{SWE}}$ is associated to the stepwise Floquet evolution constituted of $s$ steps $(U_{1}, \dots, U_{s})$, the product of which is a Floquet operator $U_{\text{F}}^{(n)}$.
We expect that such invariants are related, especially in view of the relation~\eqref{scat_power_s_block_diagonal} between $\scat^s$ and the $U_{\text{F}}^{(n)}$.

\smallbreak

Because of the structure constraint~\eqref{eq_structure_constraint}, the spectrum of $\scat$ is redundant and can be fully deduced from a fundamental domain $F$ (as illustrated in figure~\ref{relation_S_Ufn}). This property translates simply on the invariant $W_\eta^\text{HC}$, which are also partially redundant. Namely, from equation~\eqref{relation_W_chern} we have
\begin{equation}\label{W_scat_redundancy}
W_{\eta + \frac{2\pi}{s}}^\text{HC}[\scat] = W_{\eta}^\text{HC}[\scat] - C_1(P_{\eta,\eta+ \frac{2\pi}{s}}) = W_{\eta}^\text{HC}[\scat]
\end{equation} 
since the first Chern number on a fundamental domain vanishes, $C_1(P_{\eta,\eta + \frac{2\pi}{s}}) = C_1(\Pi) = 0$, see equation~\eqref{vanishing_chern}. As a consequence, $W_{\eta}^\text{HC}[\scat]$ is $2\pi/s$-periodic and the system is fully characterized by computing the $W$ invariant over a fundamental domain only.\footnote{Note that this applies for any phase rotation symmetric systems, and not only cyclic oriented networks.}

\smallbreak

We are now able to state the main result of this section, that is,
\begin{equation}\label{equality_W}
W_{\eta/s}^\text{HC}[\scat] = W_{\eta}^{\text{SWE}}[U_{\text{F}}^{(n)}]. 
\end{equation}
On the left-hand side, we know from~\eqref{W_scat_redundancy} and from the previous paragraph that the invariant is defined for $\eta/s$ without ambiguity, so that the previous formula is still $2\pi$-periodic in $\eta$. On the right hand side, we know [see~\eqref{eq:equal_SWE} and appendix~\ref{app:proofWSWE}] that the invariants associated to any $U_{\text{F}}^{(n)}$ with $n \in 1, \dots, s$ are all equal. 

The previous equality can be interpreted as follows: the topological information\footnote{In particular, note that the $W$'s also contain the first Chern numbers of the different bands.} from the one-step evolution of an oriented network ruled by $\scat \in U(\sizenode \times s)$ is fully equivalent to the one from the $s$-step stepwise Floquet evolution with steps $(U_1, \dots, U_s)$ with $U_j \in U(\sizenode)$ leading to the Floquet operator $U_{\text{F}}^{(1)} \in U(\sizenode)$ or any of its cyclic permutations $U_{\text{F}}^{(n)}$ that appear in the block diagonal operator $\scat^s$. From the topological point of view, these are two equivalent descriptions of the same problem.

\smallbreak

The identity~\eqref{equality_W} is proven by direct computation of both invariants, which are proven equal through the relation between $\scat^s$ and $U_{\text{F}}^{(n)}$.
The actual proof is quite technical and therefore postponed to appendix~\ref{app:proofW}, but we encourage the reader to have a glimpse at it.

\subsubsection{The relative nature of the invariants}
\label{sec:relative}

As we have seen in the example of the L-lattice in section~\ref{sec:L_lattice}, the invariants for cyclic oriented networks are actually \emph{relative} invariants, in a way which is very similar (and formally equivalent for $s=2$) to the standard chiral symmetric (class AIII) topological insulators, which are very clearly discussed in reference~\cite{Thiang2015b}. More precisely, such invariants are relative to \emph{reference evolution} which satisfies the structure constraint. There is indeed a large \enquote{gauge freedom} in this choice: starting from a given reference evolution $U_{\text{ref}}$, the conjugation by any change of basis matrix $M(k)$ commuting with $D$ gives another equally valid reference evolution $M(k) U_{\text{ref}} M^{-1}(k)$. The relative invariants with respect to such evolutions are indeed generally not equal. 

\smallbreak

This relative character of the topological invariants is particularly clear in the Ho-Chalker point of view, where the reference evolution is indeed chosen as the constant Bloch evolution operator
\begin{equation}\label{choice_ref_sc}
U_{\text{ref}}(t,k) =
 \begin{pmatrix}
  0 & 0 & \cdots & \Id \\
 \Id & 0 & \cdots & 0 \\
  \vdots  & \ddots  & \ddots & \vdots  \\
 0 &  \cdots & \mathcal \Id & 0 
 \end{pmatrix}
\end{equation}
satisfying the structure constraint and from which the interpolation starts. Though it may not be as obvious, the invariant defined in the Floquet point of view is indeed also relative.

\smallbreak

Notably, the choice as a reference of the Bloch evolution operator $U_{\text{ref}}(t,k)$ defined in equation~\eqref{choice_ref_sc} makes the invariants depend on the choice of the unit cell, as we have observed in the case of the L-lattice. This is because the Bloch representation as a $k$-dependent matrix $U_{\text{ref}}(t,k)$ (like in equation~\eqref{choice_ref_sc}) of the operator $U_{\text{ref}}(t)$ (acting on the Hilbert space) actually depends on the choice of the unit cell~\cite{Fruchart2014}. Importantly, the difference of topological indices at an interface and for a given gap is well defined and unambiguous.

\subsubsection{Gapless steps and their ambiguities \label{subsec:gapless}}

Let us illustrate a possible ambiguity of our definitions in the case of the L-lattice (see section~\ref{sec:L_lattice}). In classical loop configurations, at $\theta=0$ and $\theta=\pi$, we observe that the step operator $U_2(k)$ is gapless. In principle, this prevents from defining a proper topological invariant. On the other hand, this situation should only arise in critical situations when (a part of) the nodes behave classically (as perfectly reflecting or transmitting elements). Such situations could either arise at the middle of a phase or at a transition point. In the first case (an example of which is provided by the L-lattice), we expect that the limit values of the invariants on all sides of the critical manifold should agree: in this case, the common limit value can be taken as \emph{the} value of the invariant at that point. On the other hand, when the critical point marks a phase transition, we do not expect a well-defined topological invariant, and the various limits are indeed expected to disagree.


\section{Conclusion}

In this paper, we have introduced the phase rotation symmetry, a new symmetry specific to dynamical systems described by a unitary operator, which has no equivalent in Hamiltonian systems. We then  illustrated its power on our main subject, the description and the topology of oriented scattering network models. 

\smallbreak

As we have seen, the phase rotation symmetry is the signal of a redundancy in the description of the system, and allows one to reduce and simplify this description, but also to better understand the internal structure of the system. We introduced a particular class of \emph{cyclic} scattering networks where wave packets always encounter the same cycle of nodes. At the level of their evolution operator, such network models are characterized by a particular phase rotation symmetry, which is always present, and which we call a structure constraint. Two important consequences stem from the existence of this particular phase rotation symmetry. The first one is that the description of the cyclic network model can be reduced to a particular form, and then fully mapped into a stepwise Floquet dynamics. This mapping justifies and expresses the relationship between scattering networks and Floquet dynamics found in several works~\cite{Ho1996,Klesse1999,Janssen1999,Pasek2014}. A second consequence is that the structure constraint allows one to define bulk topological invariants which fully account for the topology of the network model, and in particular for topological anomalous phases. Even though such phases were already experimentally observed~\cite{Hu2015, Gao2016}, such an invariant was not known until now. Notably, the topological invariants can be defined directly from the \emph{constrained} evolution operator of the network model, or from the corresponding Floquet dynamics: both points of view indeed coincide, but they may be equally useful in different situations.

\smallbreak

Scattering network models notably allow for topological anomalous phases, where the system is topologically nontrivial despite the vanishing of all first Chern numbers. Such phases are particularly interesting, and we may want to \emph{design} them. Although the phase rotation symmetry alone is not sufficient to ensure that a phase is topologically anomalous, a strong version of this property is actually enough to ensure that all first Chern numbers vanish, which is of clear interest to the design of anomalous phases. An example of a procedure to identify anomalous phases in network models based on phase rotation symmetries in classical loop configurations is also proposed, which can also be applied to engineer anomalous Floquet phases in other kinds of systems.

\bigbreak

Whilst it was mainly used to study scattering networks, the phase rotation symmetry is a new item in the toolbox of (potentially topological) general unitary evolutions, where it can be applied both as a reduction procedure and as a design principle. Several generalizations of the phase rotation symmetry can be imagined, for example where the symmetry operator is antiunitary, i.e. where
\begin{equation}
    \rot \overline{U} \rot^{-1} = \ee^{\ii \zeta} U.
\end{equation}
It is also possible to consider more exotic generalizations: the antiunitary phase rotation constraint can be reformulated as $\rot U \rot^{-1} = \ee^{\ii \zeta} \overline{U}$ (with different $\rot$ and $\zeta$), and we can consider other constraints where $\overline{U}$ is replaced by, e.g., $U^{-1}$ (which corresponds to an actual chiral symmetry when $\zeta=0$), $U^{T}$ or $U^{\alpha}$. Some of such generalizations seem to appear in stepwise evolutions. Indeed, another very simple generalization, which should be physically relevant, consists of including the possibility of a nontrivial action on the Brillouin zone, where, for example, a constraint like $\rot U(k) \rot^{-1} = \ee^{\ii \zeta} U(-k)$ could be considered. This assortment of examples aims at highlighting that the world of \enquote{unitary} dynamical evolutions is far richer than its Hamiltonian counterpart: new kinds of effective constraints or symmetries can emerge, the phase rotation symmetry being the prime example of such. Whether such constraints deserve or not to be named \enquote{symmetries} depends on the context and on the meaning we attribute to the word. For instance, the phase rotation symmetry does indicate a redundancy in the description, which may be \enquote{broken} in other physical situations. As such, we believe that this label is indeed relevant in the context of the effective description of wave propagation. The structure constraint of cyclic scattering networks seems to \enquote{protect} the topological phase, in the same way than standard symmetries are necessary for symmetry-protected topological phases to exist.

\smallbreak

This statement can serve as an interpretation of the fact that our approach to characterize the topology of oriented scattering networks only covers the particular class of cyclic network models. Other kinds of (spatially periodic) network models exist, which can also display anomalous topological states, yet evade our characterization. A topological characterization of such systems based on the same principle should be possible, but requires further analysis.
Another open question involves the effects of a structure-constraint-breaking defects or disorder on the topological phases. Physically, such imperfections are not necessarily present in experimental realizations, but they may arise quite naturally, and we expect they should at some point spoil the topology; the question is to what extend they may be tolerated while still keeping protected edge states.

\begin{acknowledgments}
This work was supported by the French Agence Nationale de la Recherche (ANR) under Grant TopoDyn (ANR-14-ACHN-0031).
The work of C.T. was supported by the PRIN project \enquote{Mathematical problems in kinetic theory and applications} (prot. 2012AZS52J).
\end{acknowledgments}

\begin{appendices}
\makeatletter
\renewcommand{\theequation}{\thesection.\arabic{equation}}
\@addtoreset{equation}{section}
\makeatother

\section{The standard phase rotation operator \label{app:standard_PRS}}

Let us consider a phase rotation operator $\rot$ for the evolution operator $U$, such that
\begin{equation}
	\rot U \rot^{-1} = \ee^{\ii 2 \pi/M} U.
\end{equation}
In general, the phase rotation operator has no special form.
In this appendix, we show that when
\begin{itemize}
	\item the $M^{\text{th}}$ power of $\rot$ is scalar\footnote{In particular, this is necessarily the case in an irreducible representation space, where all symmetries, including $\rot^M$, are scalar.},
	that is to say $\rot^M = \ee^{\ii \phi} \, \Id$ and
	\item the evolution operator $U$ is gapped,
\end{itemize}
then the phase rotation operator $\rot$ assumes the \emph{standard form}
\begin{equation}
	\rot \simeq \rot_0 \equiv \text{diag}(1,\ee^{\ii 2\pi/M}, \ee^{\ii 4\pi/M}, \dots , \ee^{\ii 2 \pi (M-1)/M}) \otimes \Id_{\sizenode} \in U(\sizenode \times M)
\end{equation}
in an adequate basis. 
The standard phase rotation operator emphasizes the cyclic nature of the phase rotation symmetry (fundamental domains are simply rotated by the action of the operator $\rot_0$). 
When $\rot$ is a phase rotation symmetry of $U$ (namely $\rot U \rot^{-1} = \ee^{\ii 2 \pi/M} U$) and $R$ is a symmetry of $U$ (namely $R U R^{-1} = U$), then $\rot R$ and $R \rot$ are both phase rotation symmetries of $U$. Hence, many phase rotations symmetries can be constructed from the standard phase rotation operator $\rot_0$, when it is a phase rotation symmetry of $U$, which may not be reduced to the standard form. 
In general, it may also happen that the operator $\rot = \rot_0 R$ \emph{is} a phase rotation symmetry, while $\rot_0$ is \emph{not}.

\bigbreak

Let us now prove the preceding statement. First, we redefine the phase rotation operator so that $\rot^M = \Id$ by replacing $\rot$ with $\ee^{-\ii \phi/M} \rot$. 

Let then $F$ be a fundamental domain for the phase rotation symmetry, chosen to have its ends in a gap of $U$ (this is possible because we assumed that $U$ is gapped, as explained in section \ref{subsec:PRS}). Let $\psi_1, \dots, \psi_{\sizenode}$ be the eigenstates of $U$ with eigenvalue in $F$. Because of the phase rotation symmetry, the family
\begin{equation}
	(\psi_i, \dots, \psi_{\sizenode}, \rot \psi_1, \dots, \rot \psi_{\sizenode}, \dots, \rot^{M-1} \psi_1, \dots, \rot^{M-1} \psi_{\sizenode})
\end{equation}
is a basis. In this basis, $\rot$ is block-diagonal, and assumes the form
\begin{equation}
	\rot \simeq B \otimes \Id_{\sizenode}
\end{equation}
where $B$ reads
\begin{equation}
B \simeq
 \begin{pmatrix}
  0 & 0 & \cdots & 1 \\
  1 & 0 & \cdots & 0 \\
  \vdots  & \ddots  & \ddots & \vdots  \\
  0 &  \cdots & 1 & 0 
 \end{pmatrix}
\end{equation}
As $B$ is a $M \times M$ circulant matrix, it is diagonalizable and its eigenvalues are the $M^{\text{th}}$ roots of unity. Hence, $\rot$ is then diagonalized as
\begin{equation}
	 \rot \simeq \text{diag}(1,\ee^{\ii 2\pi/M}, \ee^{\ii 2\pi \times 2/M}, \dots , \ee^{\ii 2 \pi (M-1)/M})  \otimes \Id_{\sizenode}
\end{equation}
which concludes the proof.


\section{Proof of the equality between the SWE-invariants of all circular permutations $U_{\text{F}}^{(n)}$ \label{app:proofWSWE}}

We start by proving identity~\eqref{eq:equal_SWE} for $n=2$, namely that $W^{\text{SWE}}_{\eta}[U_{\text{F}}^{(1)}] = W^{\text{SWE}}_{\eta}[U_{\text{F}}^{(2)}]$, in order not to overload the explicit expressions. The generalization to any $n$ is straightforward, as discussed in the end of the paragraph. The choice of times $t_j$ in interpolation~\eqref{eq:U_interpol} is completely arbitrary and does not change the value invariant: they actually do not appear in the computation, as it can be seen in expression~\eqref{eq:explicitW_SWE}, for example. Thus, from now on, we can choose the natural and regular time-step: $t_j = jT/s$, so from the definitions~\eqref{eq:def_V} and~\eqref{def_W} the computation of $W^{\text{SWE}}_{\eta}[U_{\text{F}}^{(1)}]$ is reduced to the degree of the map
\begin{equation}
V^{(1)}_\eta(t,k) = \left\lbrace \begin{array}{lcl}
\mathcal U_{\rm int}[U_1]\big( t s \big) \ee^{\ii t  H_{\eta}^{\rm eff}[U_{\text{F}}^{(1)}]} & & 0  \leq t \leq  \dfrac{_T}{^s} \\
\mathcal U_{\rm int}[U_j]\big( (t-(j-1)\frac{_T}{^s})s \big)U_{j-1} \ldots U_1 \ee^{\ii t  H_{\eta}^{\rm eff}[U_{\text{F}}^{(1)}]} & & (j-1)\dfrac{_T}{^s} \leq t \leq j \dfrac{_T}{^s} \\
\mathcal U_{\rm int}[U_s]\big( (t-(s-1)\frac{_T}{^s})s \big)U_{s-1} \ldots U_1 \ee^{\ii t  H_{\eta}^{\rm eff}[U_{\text{F}}^{(1)}]} & & (s-1)\dfrac{_T}{^s} \leq t \leq T
\end{array}
\right.
\end{equation}
which is defined piecewise for $j\in\{1,\ldots s\}$, and where we dropped the $k$ dependency on the right hand side. Similarly, the computation of $W^{\text{SWE}}_{\eta}[U_{\text{F}}^{(2)}]$ is reduced to the degree of the map
\begin{equation}
V^{(2)}_\eta(t,k) = \left\lbrace \begin{array}{lcl}
\mathcal U_{\rm int}[U_2]\big( t s \big) \ee^{\ii t  H_{\eta}^{\rm eff}[U_{\text{F}}^{(2)}]} & & 0  \leq t \leq  \dfrac{_T}{^s} \\
\mathcal U_{\rm int}[U_{j+1}]\big( (t-(j-1)\frac{_T}{^s})s \big)U_{j} \ldots U_2 \ee^{\ii t  H_{\eta}^{\rm eff}[U_{\text{F}}^{(2)}]} & & (j-1)\dfrac{_T}{^s} \leq t \leq j \dfrac{_T}{^s} \\
\mathcal U_{\rm int}[U_1]\big( (t-(s-1)\frac{_T}{^s})s \big)U_{s} \ldots U_2 \ee^{\ii t  H_{\eta}^{\rm eff}[U_{\text{F}}^{(2)}]} & & (s-1)\dfrac{_T}{^s} \leq t \leq T
\end{array}
\right.
\end{equation}
In order to show that the degrees of these two maps are equal, we will use the homotopy invariance of the degree~\cite{Rudner2013}. Consider the following smooth deformation
\begin{equation}
\widetilde V(r;t,k) = V^{(2)}_\eta(t,k) \mathcal U_{\rm int}[U_1](rT,k) \ee^{\ii r \frac{T}{s}  H^{\rm eff}_\eta[U_{\text{F}}^{(1)}](k)}
\end{equation}
where $r \in [0,1]$ is a deformation parameter. Obviously one has $\widetilde V(0;t,k) = V^{(2)}_\eta(t,k)$.
The expression at $r=1$ is somehow close to $V^{(1)}_\eta$ since $\mathcal U_{\rm int}[U_1](T,k)=U_1$. We deduce from~\eqref{def_UFn} that
\begin{equation}
U_{\text{F}}^{(2)} = U_1  U_{\text{F}}^{(1)} U_1^{-1}
\qquad \Rightarrow \qquad
\ee^{\ii t  H_{\eta}^{\rm eff}[U_{\text{F}}^{(2)}]} = U_1 \ee^{\ii t  H_{\eta}^{\rm eff}[U_{\text{F}}^{(1)}]} U_1^{-1}
\end{equation}
which follows from the spectral decomposition of definition~\eqref{def:Heff}. Hence,
\begin{equation}
\widetilde V(1;t,k) = \left\lbrace \begin{array}{ll}
\mathcal U_{\rm int}[U_2]\big( t s \big) U_1 \ee^{\ii (t+\frac{T}{s})  H_{\eta}^{\rm eff}[U_{\text{F}}^{(1)}]}  & 0  \leq t \leq  \dfrac{_T}{^s} \\
\mathcal U_{\rm int}[U_{j+1}]\big( (t-(j-1)\frac{_T}{^s})s \big)U_{j} \ldots U_2 U_1 \ee^{\ii (t+\frac{T}{s})  H_{\eta}^{\rm eff}[U_{\text{F}}^{(1)}]}  & (j-1)\dfrac{_T}{^s} \leq t \leq j \dfrac{_T}{^s} \\
\mathcal U_{\rm int}[U_1]\big( (t-(s-1)\frac{_T}{^s})s \big)U_{s} \ldots U_2 U_1 \ee^{\ii (t+\frac{T}{s}) H_{\eta}^{\rm eff}[U_{\text{F}}^{(1)}]}  & (s-1)\dfrac{_T}{^s} \leq t \leq T
\end{array}
\right.
\end{equation}
which looks like $V_\eta^{(1)}$ but somewhat shifted in time. By homotopy invariance
\begin{equation}
W^{\text{SWE}}_{\eta}[U_{\text{F}}^{(2)}]  =  \deg\big( V_\eta^{(2)}\big) = \deg\big( \widetilde V(0;\cdot)\big) = \deg\big( \widetilde V(1;\cdot)\big)
\end{equation}
and the degree integral formula~\eqref{degreeformula} can be decomposed in pieces corresponding to the different steps:
\begin{align}
&24 \pi^2 \deg\big( \widetilde V(1;\cdot)\big) \cr
& =  \sum_{j=1}^{s-1} \int_{\BZ}  \int_{(j-1)T/s}^{jT/s}  \Big(\mathcal U_{\rm int}[U_{j+1}]\big( (t-(j-1)\frac{_T}{^s})s \big) U_j \ldots U_1 \ee^{\ii (t+\frac{T}{s})  H_{\eta}^{\rm eff}[U_{\text{F}}^{(1)}]} \Big)^*\chi \cr
& \qquad + \int_{\BZ}  \int_{(s-1)T/s}^{T}  \Big(\mathcal U_{\rm int}[U_{1}]\big( (t-(s-1)\frac{_T}{^s})s \big) \ee^{\ii (t-(s-1)\frac{T}{s})  H_{\eta}^{\rm eff}[U_{\text{F}}^{(1)}]} \Big)^*\chi 
\end{align}
where $\BZ$ is the Brillouin zone and where in the last part we have used the fact that
\begin{equation}
U_s \ldots U_1 = U_{\text{F}}^{(1)} = \ee^{-\ii T \mathcal H_\eta^{\rm eff}[U_{\text{F}}^{(1)}]}.  
\end{equation}
Then by a change of variable $t \mapsto t-T/s$ for the first term, and $t \mapsto t-(s-1)T/s$ for the second, we end up by reordering the terms as
\begin{align}
&24 \pi^2 \deg\big( \widetilde V(1;\cdot)\big) \cr
& =  \sum_{j=1}^{s} \int_{\BZ}  \int_{(j-1)T/s}^{jT/s}  \Big(\mathcal U_{\rm int}[U_{j}]\big( (t-(j-1)\frac{_T}{^s})s \big) U_{j-1} \ldots U_1 \ee^{\ii t  H_{\eta}^{\rm eff}[U_{\text{F}}^{(1)}]} \Big)^*\chi \cr
&=24 \pi^2 \deg\big( V_\eta^{(1)}\big) = 24 \pi^2 W^{\text{SWE}}_{\eta}[U_{\text{F}}^{(1)}] 
\end{align}
where the empty product $U_{j-1} \ldots U_1$ is the identity for $j=1$. This concludes the proof.\hfill $\square$

\bigbreak

The generalization to $W^{\text{SWE}}_{\eta}[U_{\text{F}}^{(1)}] = W^{\text{SWE}}_{\eta}[U_{\text{F}}^{(n)}] $ for any $n$ is straightforward by noticing that
\begin{equation}
  U_{\text{F}}^{(n)} = (U_{n-1}\ldots U_1) \, U_{\text{F}}^{(1)} \, (U_{n-1}\ldots U_1)^{-1}
\end{equation}
and by defining the corresponding homotopy
\begin{equation}
\widetilde V(r;t,k) = V^{(n)}_\eta(t,k) \mathcal U_{\rm int}[U_{n-1} \ldots U_1](rT,k) \ee^{\ii r (n-1) \frac{T}{s}  H^{\rm eff}_\eta[U_{\text{F}}^{(1)}](k)}.
\end{equation}

\section{Proof of the identity~(\ref{equality_W}) between the one-step and the stepwise invariants \label{app:proofW}}
 
The proof of the identity~\eqref{equality_W} between the one-step invariant $W_{\eta/s}^\text{HC}[\scat]$ in the Ho-Chalker point of view and the stepwise invariant $W^{\text{SWE}}_{\eta}[U_{\text{F}}^{(n)}]$ in the Floquet point of view is done by direct computation of each invariant, and using the fact that $\scat^s$ is related to the $U_{\text{F}}^{(n)}$. 

\smallbreak

First we start with the one-step evolution invariant
\begin{equation}
W_{\eta}^\text{HC}[\scat] = \frac{1}{24\pi^2} \int_{[0,T]\times \BZ} \left( \mathcal U_{\rm{int},D}[\scat] \ee^{\ii t H^{\rm eff}_\eta[\scat]} \right)^* \chi
\end{equation}
see~\eqref{def_W} and~\eqref{def_UintD_S}, where $V^*\chi = \tr( (V^{-1} \dd V)^3)$. Then, using identity 
\begin{equation}\label{eq:chiAB}
(AB)^*\chi = A^*\chi + B^\star \chi - 3 \dd \tr (A^{-1} \dd A \dd B \, B^{-1})
\end{equation}
(see e.g. appendix A of reference~\cite{Carpentier2015}), we get
\begin{equation}
W_{\eta}^\text{HC}[\scat] = \frac{1}{24\pi^2} \int_{[0,T]\times \BZ} \left( \mathcal U_{\rm{int},D}[\scat]  \right)^* \chi + \frac{1}{24\pi^2} \int_{[0,T]\times \BZ} \left(  \ee^{\ii t H^{\rm eff}_\eta[\scat]} \right)^* \chi
\end{equation}
Indeed by Stokes formula the third term is vanishing since it is reduced to an integration over the boundaries $t=0$ and $t=T$ of $[0,T]\times \BZ$. At $t=0$ the two maps are constant ($k$-independent) and at $t=T$ we get $\mathcal U_{\rm{int},\text{HC}}[\scat](T) = \scat$ whereas $\ee^{\ii T H^{\rm eff}_\eta[\scat]} = \scat^{-1}$, leading to $\tr (\scat^{-1} \dd \scat \dd (\scat^{-1}) \scat) = - \tr(\scat^{-1} \dd \scat)^2 = 0$ by antisymmetry. Note that even if the two quantities in the latter equation are not integers anymore, they will however respectively coincide with some terms coming from the computation of $W$ for the SWE. Before that the first part can already be improved by noticing that
\begin{align}
\mathcal U^{-1}_{\rm{int},\text{HC}}[\scat] \dd \mathcal U_{\rm{int},\text{HC}}[\scat] & = \text{diag} \left( \mathcal U^{-1}_{\rm{int}}[U_1] \dd \mathcal U_{\rm{int}}[U_1], \quad \ldots \quad, \,  \mathcal U^{-1}_{\rm{int}}[U_s] \dd \mathcal U_{\rm{int}}[U_s]  \right) \cr & = \text{diag} \left( \ee^{\ii t H^{\rm eff}_{-\pi}[U_1]} \dd \ee^{-\ii t H^{\rm eff}_{-\pi}[U_1]} ,\quad \ldots\quad , \,\ee^{\ii t H^{\rm eff}_{-\pi}[U_s]} \ee^{-\ii t H^{\rm eff}_{-\pi}[U_s]}  \right)
\end{align}
see~\eqref{def_UintD_S} and~\eqref{eq:def_standardHeff}, so that finally
\begin{equation}\label{eq:proof_lhs}
W_{\eta}^\text{HC}[\scat] =  \frac{1}{24\pi^2} \int_{[0,T]\times \BZ} \sum_{n=1}^s \left(\ee^{-\ii t H^{\rm eff}_{-\pi}[U_n] } \right)^* \chi + \frac{1}{24\pi^2} \int_{[0,T]\times \BZ} \left(  \ee^{\ii t H^{\rm eff}_\eta[\scat]} \right)^* \chi
\end{equation}
On the other hand, the invariant of the SWE is computed similarly: from definitions~\eqref{eq:def_W_SWE}
and~\eqref{eq:U_interpol}, separating and rescaling the time of each step $t_{j-1} \leq t \leq t_j$ in the integral by a change of variables $t'=(t-t_{j-1})/(t_j-t_{j-1})$, one has the following decomposition
\begin{align}
W_{\eta}^{\rm SWE}[U_{\text{F}}^{(1)}] = &\, \frac{1}{24\pi^2} \int_{[0,T]\times \BZ} \left( \mathcal U_{\rm{int}}[U_1]  \right)^* \chi + \left( \mathcal U_{\rm{int}}[U_2]U_1  \right)^* \chi +\dots + \left( \mathcal U_{\rm{int}}[U_s] U_{s-1} \ldots U_1 \right)^* \chi \cr & + \frac{1}{24\pi^2} \int_{[0,T]\times \BZ} \left(  \ee^{\ii t H^{\rm eff}_\eta[U_{\text{F}}^{(1)}]} \right)^* \chi
\end{align}
Then using again identity~\eqref{eq:chiAB}, the fact that $U_n^*\chi =0$ since $\chi$ is a 3-form and $U_n$ only depends on the two-dimensional variable $k$ and not on $t$, and the boundary values of $\mathcal U_{\rm{int}}[U_n] = \Id, \, U_n$ at $t=0,\, T$ respectively, we get
\begin{align}\label{eq:explicitW_SWE}
W_{\eta}^{\rm SWE}[U_{\text{F}}^{(1)}] = &\, \frac{1}{24\pi^2} \int_{[0,T]\times \BZ} \sum_{n=1}^s  \left(  \ee^{-\ii t H^{\rm eff}_{-\pi}[U_n]} \right)^* \chi  + \left(  \ee^{\ii t H^{\rm eff}_\eta[U_{\text{F}}^{(1)}]} \right)^* \chi \cr & - \frac{1}{8\pi^2} \int_{\BZ} \sum_{n=2}^s \tr \left( U_n^{-1}\dd U_n \dd(U_{n-1} \ldots U_1) (U_{n-1} \ldots U_1)^{-1} \right)
\end{align}
We see similarities between~\eqref{eq:proof_lhs} and the latter equation, however it involves the particular choice of $U_{\text{F}}^{(1)}$ whereas the first one involves $\scat$ that is in some sense more symmetric. Hence to see the equality between the two invariants we use the following trick: since all the $W_{\eta}^{\rm SWE}[U_{\text{F}}^{(j)}]$ are all equal from appendix~\ref{app:proofWSWE}, we can symmetrize the previous quantity as
\begin{align}\label{eq:proof_rhs}
W_{\eta}^{\rm SWE}[U_{\text{F}}^{(1)}]   = &  \,\dfrac{1}{s} \sum_{j=1}^s W_{\eta}^{\rm SWE}[U_{\text{F}}^{(j)}]\cr
= & \, \frac{1}{24\pi^2} \int_{[0,T]\times \BZ} \sum_{n=1}^s  \left(  \ee^{-\ii t H^{\rm eff}_{-\pi}[U_n]} \right)^* \chi  + \dfrac{1}{s} \sum_{j=1}^s \left(  \ee^{\ii t H^{\rm eff}_\eta[U_{\text{F}}^{(j)}]} \right)^* \chi \cr & - \frac{1}{8\pi^2} \int_{\BZ} \dfrac{1}{s} \left(\sum_{n=2}^s \tr \left( U_n^{-1}\dd U_n \dd(U_{n-1} \ldots U_1) (U_{n-1} \ldots U_1)^{-1} \right) + \circlearrowleft \right).
\end{align}
Indeed the first term is already symmetric and then remain unchanged, whereas the two other terms appear symmetrized, where $\circlearrowleft$ corresponds to all the possible cyclic permutations. 

\smallbreak

For example, when $s = 3$ the last term is simply equal to
\begin{align}
&\dfrac{1}{3} \left(\sum_{n=2}^3 \tr \left( U_n^{-1}\dd U_n \dd(U_{n-1} \ldots U_1) (U_{n-1} \ldots U_1)^{-1} \right) + \circlearrowleft \right) \cr
& =  \dfrac{1}{3} \Big( U_2^{-1} \dd U_2 \dd (U_1) U_1^{-1} + U_3 \dd U_3 \dd (U_2 U_1) (U_2 U_1)^{-1} \cr
 & \hspace{1cm} + U_3^{-1} \dd U_3 \dd (U_2) U_2^{-1} + U_1 \dd U_1 \dd (U_3 U_2) (U_3 U_2)^{-1}  \cr
 & \hspace{1cm} + U_1^{-1} \dd U_1 \dd (U_3) U_3^{-1} + U_2 \dd U_2 \dd (U_1 U_3) (U_1 U_3)^{-1}  \Big)
\end{align}
each line corresponding to one of the cyclic permutations of $(3,2,1)$, namely $(1,3,2)$ and $(2,1,3)$.

Coming back to the general case and comparing~\eqref{eq:proof_lhs} with~\eqref{eq:proof_rhs}, we see that the identity between the two invariants holds if and only if we have the following equality
\begin{align}\label{eq:to_prove}
&\frac{1}{24\pi^2} \int_{[0,T]\times \BZ} \left(  \ee^{\ii t H^{\rm eff}_{\eta/s}[\scat]} \right)^* \chi \cr
&=  \, \frac{1}{24\pi^2} \int_{[0,T]\times \BZ} \dfrac{1}{s} \sum_{j=1}^s \left(  \ee^{\ii t H^{\rm eff}_\eta[U_{\text{F}}^{(j)}]} \right)^* \chi \cr 
&\hspace{1cm}  - \frac{1}{8\pi^2} \int_{\BZ} \dfrac{1}{s} \left(\sum_{n=2}^s \tr \left( U_n^{-1}\dd U_n \dd(U_{n-1} \ldots U_1) (U_{n-1} \ldots U_1)^{-1} \right) + \circlearrowleft \right)
\end{align}
Note the difference of parameters in the effective Hamiltonians, coming from~\eqref{equality_W}. This equality will be proved using the spectral decompositions of $\scat$ and $\scat^s$. The spectral decomposition of $\scat$ is
\begin{equation}
\label{}
  \scat = \sum_{r=0}^{s-1} \sum_{j=1}^{\sizenode} \ee^{-\ii 2 \pi r/s} \lambda_j  D^{r} \ket{\psi_j}\bra{\psi_j} D^{-r}
\end{equation}
due to the structure constraint~\eqref{eq_structure_constraint}, see equation \eqref{eq:spec_scat}.
Hence,
\begin{equation}
H_\eta^{\rm eff}[\scat] = 
\ii \sum_{r=0}^{s-1} \sum_{j=1}^{\sizenode} \ln_{-\eta}(\lambda_j) D^r \ket{\psi_j}\bra{\psi_j} D^{-r}
+ 2 \pi \sum_{r=1}^{s-1} \sum_{j=1}^{\sizenode} \frac{r}{s} D^r \ket{\psi_j}\bra{\psi_j} D^{-r}
\end{equation}
Besides, the $s$-th power of $\scat$ reads
\begin{equation}
\scat^{s} = \sum_{r=0}^{s-1} \sum_{j=1}^{\sizenode} \lambda_j^{s}  D^{r} \ket{\psi_j}\bra{\psi_j} D^{-r}
\end{equation}
so its effective Hamiltonian is
\begin{align}
H_\eta^{\rm eff}[\scat^s] = &\, \ii  \sum_{r=0}^{s-1} \sum_{j=1}^{\sizenode} s \ln_{-\eta}(\lambda_j) D^r \ket{\psi_j}\bra{\psi_j} D^{-r}
\end{align}
The two effective Hamiltonians are indeed not equal for the same branch cut. However, if $\lambda^s = \ee^{\ii \varphi}$ with $-\eta - 2 \pi < \varphi < -\eta$, then
\begin{equation}
-\frac{\eta}{s} - 2 \pi < -\frac{\eta}{s} - \frac{2\pi }{s} < \frac{\varphi}{s} < - \frac{\eta}{s}
\end{equation}
from which we deduce (using the definition \eqref{eq_def_log} of the logarithm) that
\begin{equation}
H_\eta^{\rm eff}[\scat^s] = s H_{\eta/s}^{\rm eff}[\scat] + 2 \pi \sum_{r=1}^{s-1} \sum_{j=1}^{\sizenode} r D^r \ket{\psi_j}\bra{\psi_j} D^{-r}
\end{equation}
On top of that, since $\scat^s$ is block diagonal, we immediately deduce its effective Hamiltonian in terms of the $U_{\text{F}}^{(n)}$ from~\eqref{scat_power_s_block_diagonal}, so we finally get
\begin{equation}
\text{diag}\left(  H_\eta^{\rm eff}[U_{\text{F}}^{(1)}], \ldots, H_\eta^{\rm eff}[U_{\text{F}}^{(s)}] \right) 
= s H_{\eta/s}^{\rm eff}[\scat] + 2 \pi \sum_{r=1}^{s-1} \sum_{j=1}^{\sizenode} r D^r \ket{\psi_j}\bra{\psi_j} D^{-r}
\end{equation}
We now wish to take the exponential of $\ii t$ times this equality, and to compute the 3-form $\chi$ on the result. The two terms on the right hand side commute because they are both decomposed on the mutually orthogonal projectors $D^{r} \ket{\psi_j}\bra{\psi_j} D^{-r}$, and the left hand side is block diagonal so
\begin{equation}
\sum_{n=1}^s \left( \ee^{\ii t H_\eta^{\rm eff}[U_{\text{F}}^{(n)}] }\right)^*\chi =  \left( \Big(\ee^{\ii t H_{\eta/s}^{\rm eff}[\scat]  }\Big)^s \prod_{r=1}^{s-1} D^r \ee^{\ii 2 \pi r t \Pi}D^{-r}\right)^*\chi
\end{equation}
where $\Pi \equiv \ket{\psi_1}\!\bra{\psi_1} + \cdots + \ket{\psi_{\sizenode}}\!\bra{\psi_{\sizenode}}$ is the projector on the fundamental domain $F$ of $\scat$ as explained in section~\ref{subsec:PRS}. Using again identity~\eqref{eq:chiAB} we get
\begin{align}
&\frac{1}{24\pi^2} \int_{[0,T]\times \BZ} \left( \Big(\ee^{\ii t H_{\eta/s}^{\rm eff}[\scat]  }\Big)^s \prod_{r=1}^{s-1} D^r  \ee^{\ii 2 \pi r t \Pi   }D^{-r}\right)^*\chi\cr
& = \frac{1}{24\pi^2} \int_{[0,T]\times \BZ}\left( \Big(\ee^{\ii t H_{\eta/s}^{\rm eff}[\scat]  }\Big)^s\right)^*\chi +  \sum_{r=1}^{s-1} \left( D^r  \ee^{\ii 2 \pi r t \Pi   }D^{-r}\right)^*\chi + 0
\end{align}
where the $0$ comes from the fact that $\ee^{-\ii 2 \pi r t \Pi   } = \Id$ both at $t=0$ and $1$. 
First, using identity (A13) of~\cite{Fruchart2016} and equation~\eqref{vanishing_chern}, we have
\begin{equation}
 \frac{1}{24\pi^2} \int_{[0,T]\times \BZ} \left( D^r  \ee^{\ii 2 \pi  r t \Pi }D^{-r}\right)^*\chi  = - r \, C_1(D^r \Pi D^{-r}) = - r \, C_1(\Pi) = 0
\end{equation}
for every $r=1,\ldots,s-1$, so that 
\begin{equation}
\frac{1}{24\pi^2} \int_{[0,T]\times \BZ} \sum_{n=1}^s \left( \ee^{\ii t H_\eta^{\rm eff}[U_{\text{F}}^{(n)}] }\right)^*\chi  = \frac{1}{24\pi^2} \int_{[0,T]\times \BZ}\left( \Big(\ee^{\ii t H_{\eta/s}^{\rm eff}[\scat]  }\Big)^s\right)^*\chi 
\end{equation}
Then, by induction on $s$ of identity~\eqref{eq:chiAB}, and with the fact that $\ee^{\ii t H_{\eta/s}^{\rm eff}[\scat]  } = \scat^{-1}$ at $t=T$ we get
\begin{align}
&\frac{1}{24\pi^2} \int_{[0,T]\times \BZ}\left( \Big(\ee^{\ii t H_{\eta/s}^{\rm eff}[\scat]  }\Big)^s\right)^*\chi \cr  & = s \frac{1}{24\pi^2} \int_{[0,T]\times \BZ}\left( \ee^{\ii t H_{\eta/s}^{\rm eff}[\scat]  }\right)^*\chi  + \frac{1}{8\pi^2} \int_{\BZ} \sum_{k=1}^{s-1} \tr\Big(\scat^{-1}\dd \scat \dd(\scat^k) \scat^{-k} \big)
\end{align}
Finally, because of the specific form of $\scat$ given by~\eqref{eq:S}, $\dd(\scat^k) \scat^{-k}$ is always block-diagonal for any $k$, with blocks of the form $\dd(U_n \ldots U_{n-k+1}) (U_n \ldots U_{n-k+1})^{-1}$ and all the corresponding cyclic permutations. From which we infer 
\begin{equation}
 \sum_{k=1}^{s-1} \tr\Big(\scat^{-1}\dd \scat \dd(\scat^k) \scat^{-k} \big) =  \left(\sum_{n=2}^s \tr \left( U_n^{-1}\dd U_n \dd(U_{n-1} \ldots U_1) (U_{n-1} \ldots U_1)^{-1} \right) + \circlearrowleft \right)
\end{equation}
Putting all together the last three equations, we get
\begin{align}
\frac{1}{24\pi^2}& \int_{[0,T]\times \BZ} \sum_{n=1}^s \left( \ee^{\ii t H_\eta^{\rm eff}[U_{\text{F}}^{(n)}] }\right)^*\chi \cr
&= s \frac{1}{24\pi^2} \int_{[0,T]\times \BZ}\left( \ee^{\ii t H_{\eta/s}^{\rm eff}[\scat]  }\right)^*\chi  \cr
&\hspace{0.5cm}+  \frac{1}{8\pi^2} \int_{\BZ} \left(\sum_{n=2}^s \tr \left( U_n^{-1}\dd U_n \dd(U_{n-1} \ldots U_1) (U_{n-1} \ldots U_1)^{-1} \right) + \circlearrowleft \right)
\end{align}
which establishes the equality~\eqref{eq:to_prove} and completes the proof of identity~\eqref{equality_W} between the two invariants.\hfill $\square$

\end{appendices}

\bibliography{bibliography}

\end{document}